\documentclass[%
,aps%
 ,reprint%
 ,secnumarabic%
,amssymb, amsmath,nobibnotes, aps, prl, floatfix]{revtex4-1}
\usepackage{docs}%
\usepackage{bm}%
\usepackage[colorlinks=true,linkcolor=blue]{hyperref}%
\expandafter\ifx\csname package@font\endcsname\relax\else
 \expandafter\expandafter
 \expandafter\usepackage
 \expandafter\expandafter
 \expandafter{\csname package@font\endcsname}%
\fi
\usepackage{graphicx}
\usepackage{array}

\begin{document}

\title{SIS epidemics on Triadic Random Graphs}%
\author{Ilja Rausch}%
\email{ilja.rausch@stud-mail.uni-wuerzburg.de}
\affiliation{Universit\"at W\"urzburg, Fakult\"at f\"ur Physik und Astronomie, Am Hubland, 97074 W\"urzburg, Germany}
\date{\today}%
\begin{abstract}
It has been shown in the past that many real-world networks exhibit community structures and non-trivial clustering which comes with the occurrence of a notable number of triangular connections. Yet the influence of such connection patterns on the dynamics of disease transmission is not fully understood. In order to study their role in the context of Susceptible-Infected-Susceptible (SIS) epidemics we use the Triadic Random Graph (TRG) model to construct small networks (N=49) from distinct, closed, directed triadic subpatterns. We compare various global properties of TRGs and use the N-intertwined mean-field approximation (NIMFA) model to perform numerical simulations of SIS-dynamics on TRGs. The results show that the infection spread on undirected TRGs displays very similar behavior to TRGs with an abundance of (directed) feed-back-loops, while using (directed) feed-forward-loops as network-entities significantly slows down the epidemic and lowers the number of infected individuals in the endemic state. Moreover, we introduce a novel stochastic approach for modelling the SIS-epidemics on TRGs based on characterizing nodes according to their set of $\left(k_{in},k_{out}\right)$ within triads. Within this model, the topology of the network is given by the number and the local structure of directed triadic motifs and not by the adjacency matrix. Nevertheless, the outcome of simulations is qualitatively similar to the results of the NIMFA model. 
\end{abstract}
\maketitle
\clearpage

\section{Introduction}
Being in contact with other people is an essential part of human life. In the framework of network analysis this phenomenon can be examined from a mathematical point of view. Networks can be found in a vast range of different fields \cite{Newman2001, Girvan, Conyon, Protein} and are thus an important part of our society. Furthermore, the interconnection of people can be treated scientifically from the perspective of epidemiology. It is obvious that the spread of many airborne and direct contact diseases is influenced by the underlying structure of interpersonal connections \cite{Klovdahl, Mossong, Colizza}. The combination of epidemic spreading and network analysis, as well as deriving corresponding mathematical models is an ongoing scientific challenge \cite{DiseaseDynamics, Danon2011, Miller2014, Satorras2015}.\\
One thorny question that arises is: how exactly does the structure of a network influence the transmission of a disease? In search for an answer to this question, many different approaches have been suggested, such as consideration of weighted networks \cite{Viboud}, dynamic networks \cite{Vernon}, generation of networks from real data \cite{Chlamydia}, random and scale-free networks \cite{Ball, Britton, Pastor}. Much attention is devoted to including realistic topological phenomena, for example the existence of community structures (or households) which lead to non-trivial clustering \cite{Girvan2004, Ball, Household}. Such community structures often display a high number of triangular connections, highlighting the importance of accounting for triangles and triadic subgraphs \cite{fagiolo2007, Ahnert}. Moreover, the abundance of loops and specific motifs have been found in numerous real-world networks \cite{Alon, Inherent, Squartini}. In this paper we examine the role of such substructures on the transmission of diseases. For this purpose we have constructed ensembles of networks which consist of directed \textit{triadic motifs} \cite{Alon} as building blocks applying the concept of \textit{Triadic Random Graphs} (TRGs) \cite{Marco}. Most publications concerned with analysis of epidemic dynamics on complex networks focus on undirected networks \cite{Danon2011, Satorras2015, Miller2014}, however numerous real-world networks are either semi-directed or fully directed \cite{epThreshold, Satorras2015}. Although challenging, including the directionality to our analysis is crucial due to the directional structure of triadic motifs. Moreover, many stochastic approaches to network epidemics consider large networks with a large or infinite number of nodes $N$ \cite{House2009,Noel2014,Mieghem2014,Miller2014}. However, on a large scale the contribution of local structure of certain motifs may become neglectable \cite{NewmanClustering}, whereas on a smaller scale their contribution to the dynamics of the system might be more apparent, which is why small networks, in our case with $N=49$, might be more appropriate.\\
In this paper we will firstly introduce the concept of Triadic Random Graphs (Sec.$\,$\ref{sec:TRG}). An extensive set of measurements performed on these graphs will be presented and used for discussions in subsequent sections. We will focus mainly on global properties that are useful in the context of epidemic dynamics. It should be emphasized that whenever TRGs are mentioned, we mean an ensemble of more than 200 samples per type of TRG over which the measured properties and simulations are averaged. However, the samples only differ in the (random) choice of \textit{Steiner Triples} and not the detailed type of their building blocks, the triadic motifs. In Sec.$\,$\ref{sec:MeanField} we address the question of how the spread of diseases behaves on TRGs with $N=49$. For this purpose different types of TRGs and their corresponding randomizations were generated. Numerical mean-field simulations of the \textit{Susceptible-Infected-Susceptible} model will be presented. We observe that for most types of triadic subgraphs the fraction of infected individuals $x$ changes with time in a very similar way. However, TRGs with an abundance of FFL-motifs ($\mathcal{T}_{1}$) show a qualitatively different behavior.\\
Furthermore, in order to get a better understanding of epidemic dynamics on TRGs and the influence of triadic motifs, it is worthwhile to consider the stochastic behavior (Sec.$\,$\ref{sec:stoch}). After a rough summary of few existing methods for regular and Erd\H{o}s-R\'enyi graphs, we will introduce a novel stochastic model for the epidemic spread on TRGs based on the assignment of characteristic \textit{tags} to the nodes. This node-tagging approach needs much less input than the method used in Sec.$\,$\ref{sec:MeanField} because it focuses only on the type and number of triadic motifs instead of including the whole topology of a preconstructed network. Remarkably, the numerical results are nonetheless qualitatively similar to those obtained in Sec.$\,$\ref{sec:MeanField}.

\section{\label{sec:TRG}Triadic Random Graphs}
\subsection{Triadic Motifs}
\label{subsec:triadicMotifs}
\begin{figure}
 \includegraphics[width=0.8\linewidth]{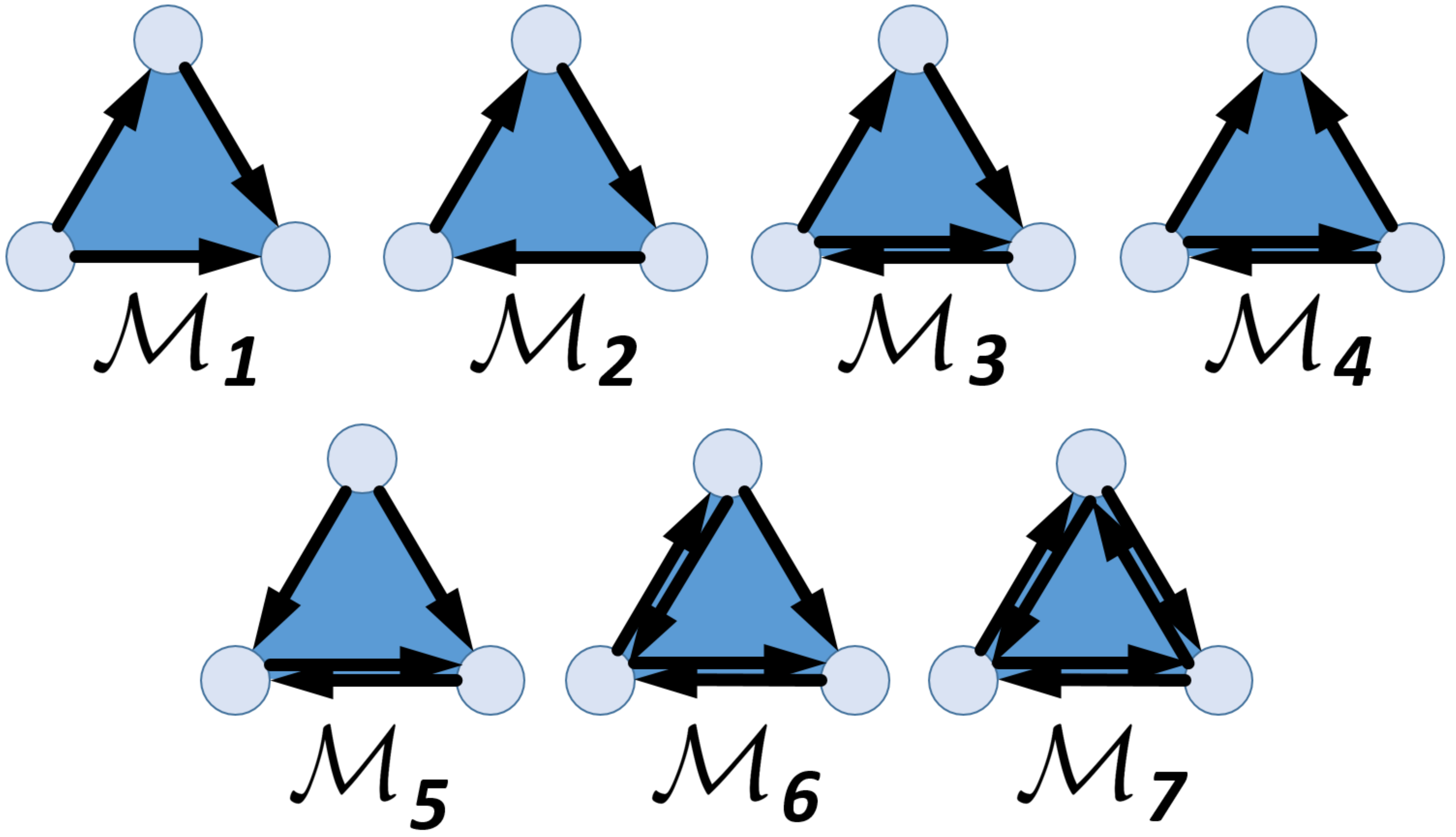}
 \caption{\label{fig:motifs}The seven types of triadic motifs which serve as building blocks for those Triadic Random Graphs $\mathcal{T}_{i}$ we are concerned with in this article.}
\end{figure}

One of the most common models for construction of random graphs is the Erd\H{o}s-R\'enyi (ER) model \cite{Erdos, Erdos60}. It describes binomial, undirected networks where the number of neighbors, i.e.~the total degree $k_{tot}$, is assigned to every node randomly. Hence, for small networks the probability for a node to have the degree $k_{tot}$ follows the Bernoulli distribution, which converges to Poisson distribution for large networks.\\
Although ER networks are easy to generate and offer insights into fully random interconnected systems, they may be too simplistic for other scientific purposes and in most cases they fail to include important features of real world networks \cite{Barabasi}, some of which are clustering, heterogeneous population structures and occurence of community structures \cite{Serrano2006, fagiolo2007, Karrer2011, Marco}. Furthermore, unlike in ER graphs, in many directed real world networks certain subgraph patterns appear evidently more often than it is expected from randomized equivalents. This was first found by R. Milo et al. \cite{Alon} who analysed a number of technological and biological data sets and detected certain patterns with above average occurrence frequency which they defined as \textit{network motifs}.\\
In the following we will focus on a particular kind of directed motifs, the (closed) \textit{triadic motif} $\mathcal{M}_{i}$ of type $i$ shown in Fig.$\,$\ref{fig:motifs}.

\subsection{Steiner Triple Systems}
\label{subsec:STS}
An interesting question emerging is: how do triadic motifs influence dynamical processes \cite{Alon2007}? In order to answer this question, it is useful to generate and compare networks which differ from each other only in the type of triadic motifs, which play the role of building blocks. So far, notable effort was devoted to finding a successful method of constructing networks with an abundance of triangles and triads \cite{Holme2002, Serrano2005, NewmanClustering, Karrer2011}. Recently, one particular model which allows to generate and analyse directed graphs with triadic motifs as primary building blocks has been presented by M. Winkler and J. Reichardt \cite{Marco}. In their publications the authors made use of the \textit{Steiner Triple System} (STS), which is a mathematical system that consists of distinct subsets of three elements (= triples). An important characteristic of the STS is that every pair of elements is connected through exactly one unique link, so that every link can be assigned to exactly one triad. Thus, a graph which is based on Steiner Triples (STs) consists of independent triadic subgraphs. In order to construct a STS with $N$ elements, two necessary and sufficient conditions have to be met \cite{Kirkman}:
\begin{equation}
\begin{aligned}
N \text{mod}\,2 &= 1 \\
N(N-1) \text{mod}\, 3 &= 0
\end{aligned}
\end{equation}
Moreover, there is an upper bound for the number of dyads-disjoint triads \cite{Marco}:
\begin{equation}
T\,\leq\,\frac{1}{3}\,=\,\frac{1}{3}\frac{N(N-1)}{2},
\end{equation}
where $M$ is the maximum number of links in a graph.

\subsection{Triadic Random Graphs}
\label{subsec:TRG}
In analogy to the conditionally independent edges in the random graphs generated by P. Erd\H{o}s and A. R\'enyi, the model introduced by M. Winkler and J. Reichardt \cite{Marco} assumes conditionally independent directed triads and the graphs generated by this model are termed \textit{Triadic Random Graphs} (TRGs). Similar to the ER model, an ensemble of random networks with the same number of nodes and edges can be generated. However, instead of randomly assigning degrees to the nodes, in the TRG model STs are assigned randomly. Hence, by choosing the type and number of triadic motifs displayed in Fig.$\,$\ref{fig:motifs} and assigning them randomly to the nodes, it is possible to construct directed networks with fixed quantities of certain overrepresented triadic motifs and random properties otherwise.\\
Additionally, to examine their influence on the dynamics of a system, it is necessary to compare TRGs to their randomized counterparts, the \textit{null-models}. The randomization is realized by a Markov Chain Monte Carlo rewiring algorithm \cite{Nospam}. Starting with a TRG sample which is to be randomized, the algorithm performs random but degree-preserving switching of edges between any pair of nodes. Consequently, the triadic structure is lost and ideally there are no overrepresented triadic motifs while the number of nodes, edges and, on average, each node's In- and Out-degrees are preserved. For more detailed information on the graph randomization, see, e.g. \cite{Nospam}. \\
In the present work, TRGs are used to analyse the impact of triadic motifs on a particular kind of dynamics: the epidemic spread of diseases. For this purpose we constructed TRGs purely from conditionally independent triadic motifs of certain kind. A TRG which consists only of triadic motifs of type $\mathcal{M}_{i}$ will be from now on denoted as $\mathcal{T}_{i}$. As mentioned before, every $\mathcal{T}_{i}$ represents an ensemble of 200 instances of the same kind except for the random distribution of STs. The total degree of every node in such a network has an even value since every node is an element of $\theta$ triads ($\theta\in\mathbb{N}$). TRGs are constructed in a way that an arbitrary node $\nu$ has the same $\theta$ in every $\mathcal{T}_{i}$ per instance. In other words, for all first instances $\theta_{\mathcal{T}_{i},\nu} = \theta_{\mathcal{T}_{j},\nu}$. The number of distinct triadic motifs in the resulting TRGs was confirmed using the MFINDER software (version 1.2) \cite{mfinder}.\\
The beauty of TRGs lies in the fact that their global properties, influenced by the abundance of $\mathcal{M}_{i}$, often show convenient similarities. Before discussing the results of performed numerical simulations of epidemic spread, it is of importance to look into some of these properties and evaluate their potential impact on the system's dynamics.

\begin{table*}
\caption{\label{tab:properties} Numerical estimates or various properties of TRGs $\mathcal{T}_{i}$ which are constructed purely from $T=60$ closed triadic motifs of type $\mathcal{M}_{i}$, averaged over ensembles of $>$200 samples. Mean values of: total degree $\left\langle k_{tot}\right\rangle$, clustering coefficient $\left\langle C\right\rangle$, diameter $\mathcal{D}$, distance $\left\langle \delta\right\rangle$, betweenness centrality $\left\langle B\right\rangle$, graph spectrum $\lambda$, assortativity $\rho$ and the current parameter $\xi$. The brackets $\left\langle { }\right\rangle$ indicate that the values were averaged over all nodes per instance of $\mathcal{T}_{i}$. The standard deviation value of $k_{tot}$ is several orders of magnitude smaller than the expectation value and can be therefore neglected. The results in the brackets show $\left\langle \delta\right\rangle$ when the infinite values are omitted. }
\begin{ruledtabular}
\begin{tabular}{c|c|c|c|c|c|c|c|c}
$\mathcal{M}_{i}$ & $\left\langle k_{tot}\right\rangle$ & $\left\langle C\right\rangle$ & $\mathcal{D}$ & $\left\langle \delta\right\rangle$ & $\left\langle B\right\rangle$ & $\lambda$ & $\rho$ & $\xi$ \\
\hline
\begin{minipage}{.05\textwidth}\includegraphics[width=\linewidth]{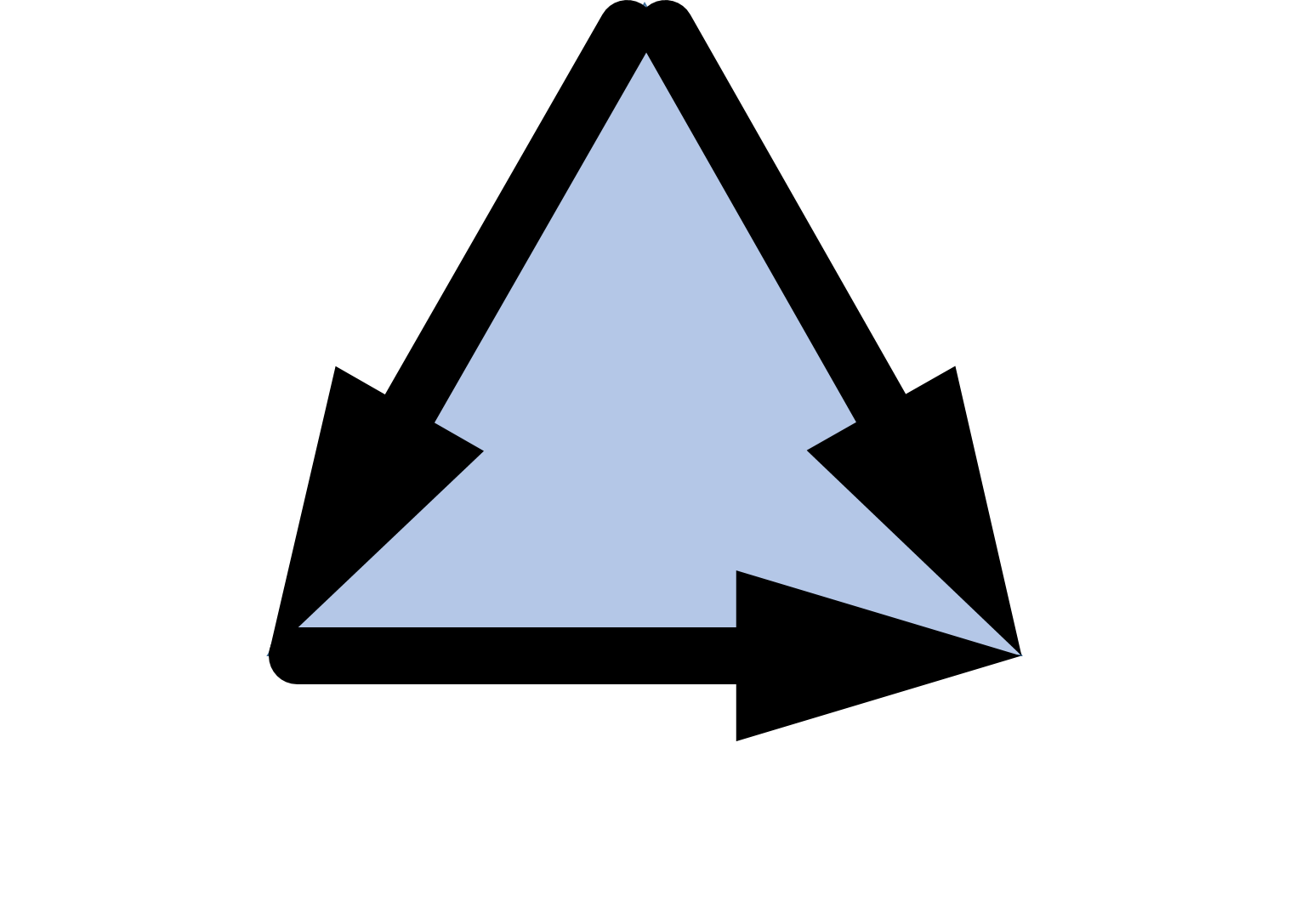}\end{minipage}  & 7.35 & 0.176 $\pm$ 0.016 & $\infty$ & $\infty \left[2.52 \pm 0.80\right]$ & 29 $\pm$ 14 & 1.21 $\pm$ 0.81 & -0.132 $\pm$ 0.085 & 0.885 $\pm$ 0.037\\
\begin{minipage}{.05\textwidth}\includegraphics[width=\linewidth]{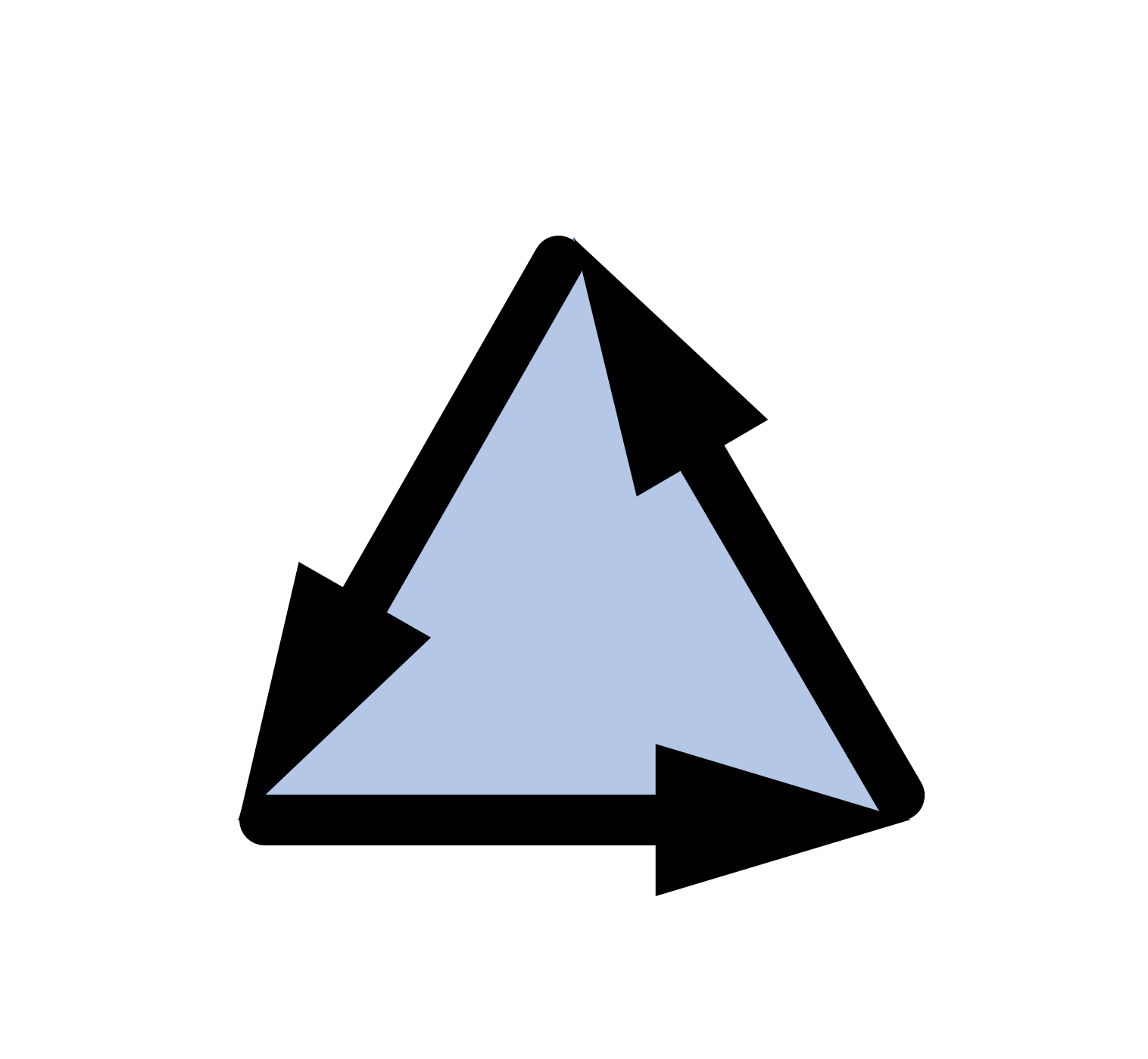}\end{minipage}  & 7.35 & 0.176 $\pm$ 0.016 & 5.73 $\pm$ 0.53 & 2.82 $\pm$ 0.35 & 87.2 $\pm$ 1.4 & 4.41 $\pm$ 0.14 & 0.53 $\pm$ 0.14 & 0.557 $\pm$ 0.020\\
\begin{minipage}{.05\textwidth}\includegraphics[width=\linewidth]{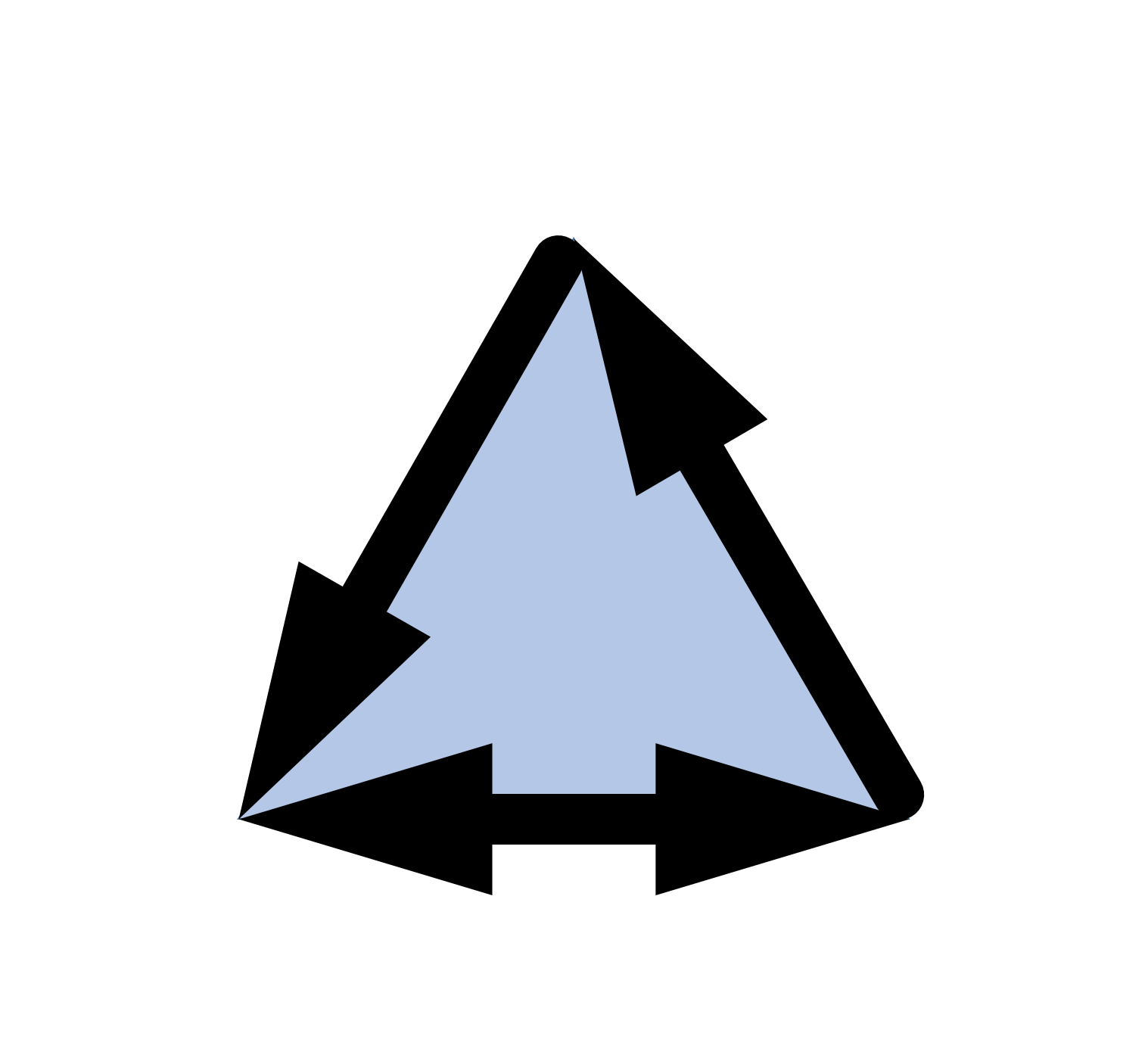}\end{minipage}  & 9.80 & 0.233 $\pm$ 0.029 & 5.21 $\pm$ 0.44 & 2.56 $\pm$ 0.33 & 75.0 $\pm$ 1.4 & 5.88 $\pm$ 0.23 & 0.43 $\pm$ 0.14 & 0.517 $\pm$ 0.018\\
\begin{minipage}{.05\textwidth}\includegraphics[width=\linewidth]{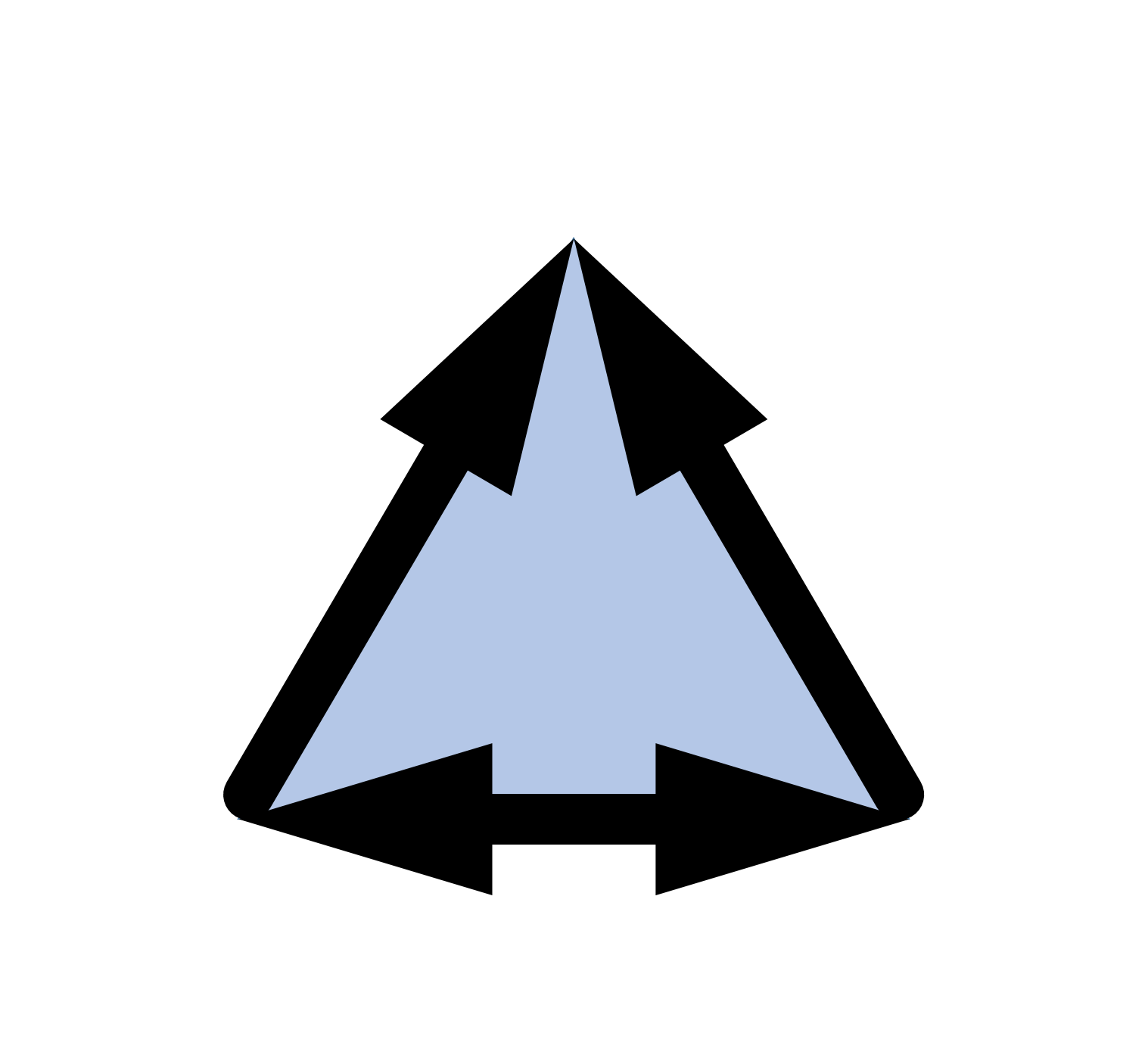}\end{minipage}  & 9.80 & 0.233 $\pm$ 0.029 & $\infty$ & $\infty \left[2.60 \pm 0.52\right]$ & 68.1 $\pm$ 7.1 & 5.25 $\pm$ 0.29  & 0.20 $\pm$ 0.12 & 0.515 $\pm$ 0.035\\
\begin{minipage}{.05\textwidth}\includegraphics[width=\linewidth]{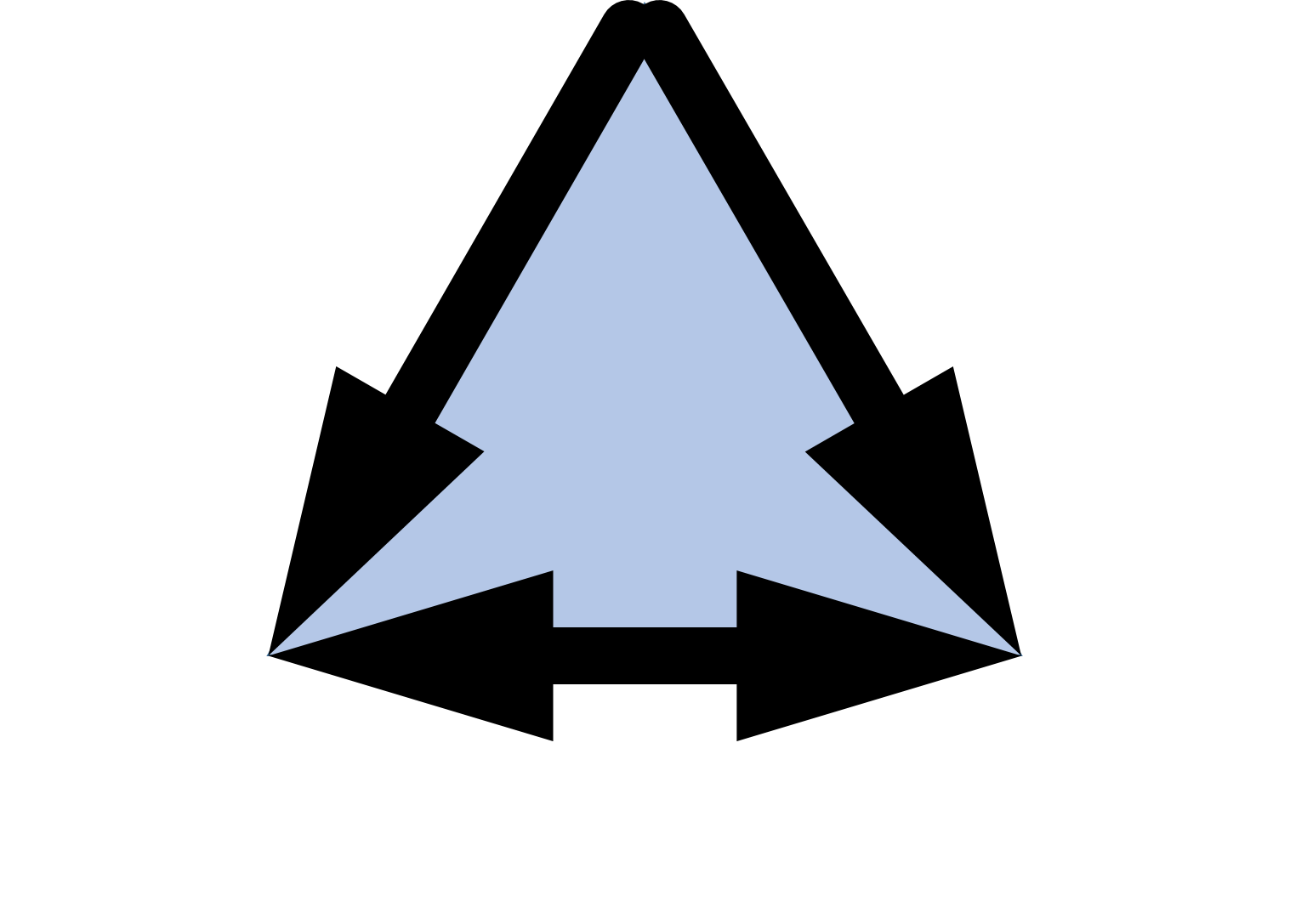}\end{minipage}  & 9.80 & 0.233 $\pm$ 0.029 & $\infty$ & $\infty \left[2.63 \pm0.40\right]$  & 68.1 $\pm$ 7.1 & 5.25 $\pm$ 0.29 & 0.20 $\pm$ 0.12 & 0.628 $\pm$ 0.067\\
\begin{minipage}{.05\textwidth}\includegraphics[width=\linewidth]{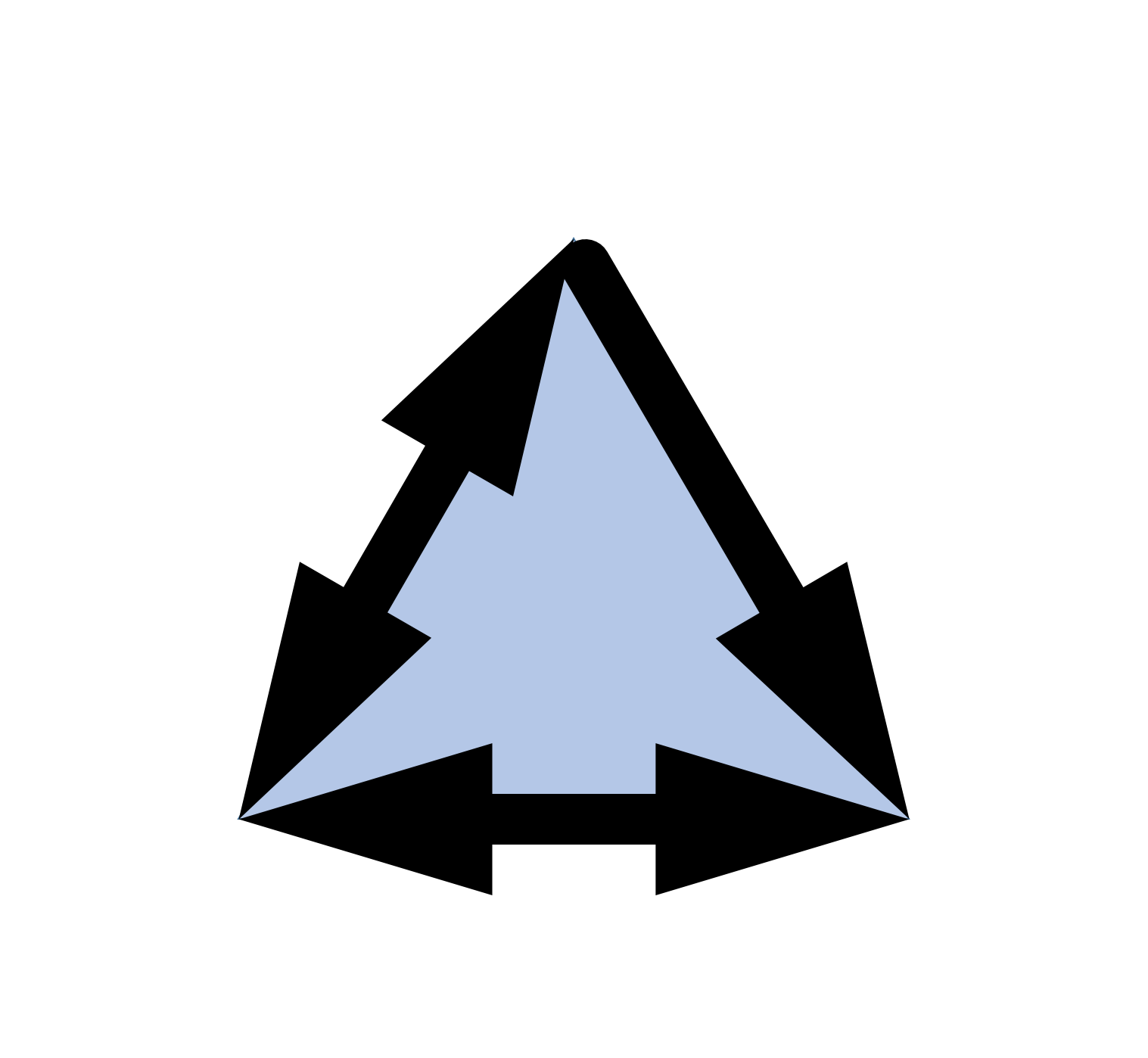}\end{minipage}  & 12.24 & 0.279 $\pm$ 0.031 & 4.77 $\pm$ 0.47 & 2.36 $\pm$ 0.29 & 65.1 $\pm$ 1.2 & 7.21 $\pm$ 0.24 & 0.38 $\pm$ 0.12 & 0.500 $\pm$ 0.017\\
\begin{minipage}{.05\textwidth}\includegraphics[width=\linewidth]{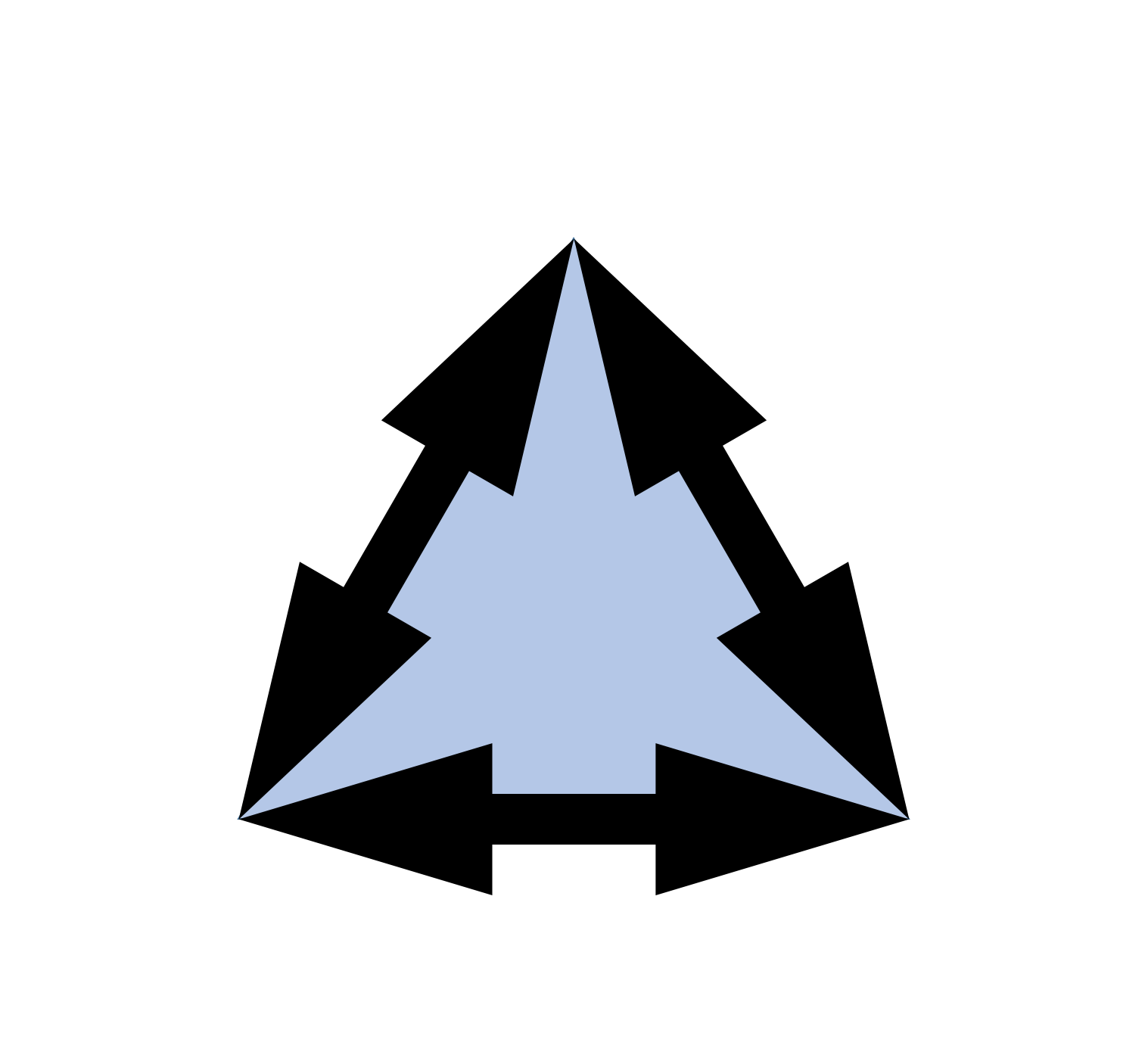}\end{minipage}  & 14.69 & 0.352 $\pm$ 0.033 & 4.11 $\pm$ 0.31 & 2.17 $\pm$ 0.26 & 56.3 $\pm$ 1.1 & 8.93 $\pm$ 0.29 & 0.53 $\pm$ 0.14 & 0.500 $\pm$ 0.001\\
\end{tabular}
\end{ruledtabular}
\end{table*}

\subsection{Degree}
\label{subsec:degreeDistribution}
As TRGs are constructed from STSs, the total degree of each node is even. To every node $\nu$ a pair of other nodes is assigned randomly and conditionally independent from other node-pairs, provided that the three nodes are elements of the same ST. Hence, the number of triadic motifs around $\nu$ is binomially distributed for small networks and Poisson distributed for large networks. However, the absolute value of the total degree is correlated with the type of the triadic motif. For instance, let us assume that $\nu$ is an element of $\theta$ triadic motifs and that all motifs are of the type $\mathcal{M}_{7}$ (see Fig.$\,$\ref{fig:motifs}). Then $\nu$'s total degree is $k_{tot,7} = 2 \theta$. However, in the case that all motifs are of the type $\mathcal{M}_{2}$, $\nu$'s total degree is $k_{tot,2} = \theta = \frac{1}{2}k_{tot,7}$. The total degree of a node $\nu$ can be calculated from the adjacency matrix $\boldsymbol{A}$: $k_{tot, \nu} = k_{in, \nu} + k_{out, \nu} = \Sigma_{j=1}^N A_{\nu j} + \Sigma_{j=1}^N A_{j\nu}$. The adjacency matrix of $\mathcal{T}_{7}$ has exactly two times more entries than the adjacency matrix of $\mathcal{T}_{1}$ or $\mathcal{T}_{2}$. 
Thus, for TRGs which differ only in choice of $\mathcal{M}_{i}$ we obtain the relations for the average degrees (averaged over all nodes of the network):
\begin{equation}
\begin{aligned}
&k_{in, 2} &=&\,\,\,k_{out, 2}=\frac{1}{2} k_{in, 7} \\
&k_{tot, 4} &=&\,\,\,k_{tot, 1} +k_{bid, 4} \\
&k_{tot, 6} &=&\,\,\,k_{tot, 1} +2\,k_{bid, 6} \\
&k_{tot, 7} &=&\,\,\,k_{tot, 1} +3\,k_{bid, 7} = 2\,k_{tot, 1},
\end{aligned}
\end{equation}
where $k_{in}$, $k_{out}$ and $k_{bid} = \left(\boldsymbol{A}^2\right)_{\nu\nu}$ denote the in-, out- and bidirected (or reciprocal) degree, respectively.
Furthermore, we find the general formulas for the average total degree $\left\langle k_{tot,i}\right\rangle$ and the number of edges $E_i$:
\begin{eqnarray}
\left\langle k_{tot,i}\right\rangle &=& 2\frac{T}{N} E(\mathcal{M}_{i}) \\*
\left\langle k_{tot,i}\right\rangle &=& \frac{E(\mathcal{M}_{i})}{E(\mathcal{M}_{j})}\left\langle k_{tot,j}\right\rangle \\*
E_i &=& \frac{E(\mathcal{M}_{i})}{E(\mathcal{M}_{j})}E_j \label{eq:edges}
\end{eqnarray}
where $E(\mathcal{M}_{i})$ denotes the number of edges within a single triadic motif of type $\mathcal{M}_{i}$ and $E_i$ the total number of edges within $\mathcal{T}_{i}$.
Some TRGs have nodes with $k_{in} = 0$ or $k_{out} = 0$, specifically $\mathcal{T}_{1}$, $\mathcal{T}_{4}$ and $\mathcal{T}_{5}$, while other $\mathcal{T}_{i}$ have no such nodes. Denoting $z_{in,i}$ ($z_{out,i}$) the number of nodes with zero in-degree (out-degree) of $\mathcal{T}_{i}$, we find that $z_{in,1} = 6.6 \pm 2.1$, $z_{in,5} = 5.4 \pm 2.3$ and zero else, and $z_{out,1} = 5.4 \pm 2.3$ and $z_{out,4} = 5.4 \pm 2.3$ and zero else.\\
The degree distribution depends on the distribution of the triadic motifs in the network and can be calculated from the TRG model. For a more detailed discussion we refer to \cite{Marco}.\\

\subsection{Clustering Coefficient}
\label{subsec:clusteringCoefficient}
The clustering coefficient aims to capture quantitatively the fraction of neighbors of a node $\nu$ which are also neighbors to each other. Communities are likely to have a high clustering coefficient, whereas it is usually low or equal to zero for random graphs such as the ER graphs.
For directed graphs with the adjacency matrix $\boldsymbol{A}$ the clustering coefficient of node $\nu$ is given by \cite{fagiolo2007}:
\begin{equation}
C_{\nu}(\boldsymbol{A}) = \frac{\theta_{\nu}}{T_{\nu}} = \frac{\left[(\boldsymbol{A}+\boldsymbol{A}^T)^3\right]_{\nu \nu}}{2\,k_{tot, \nu}(k_{tot, \nu}-1)-2\,k_{bid, \nu}},
\end{equation}
where $\theta_{\nu}$ is the number of triangles containing $\nu$ and $T_{\nu}$ is the maximum number of triangles that \textit{could} contain $\nu$. Note that $T_{\nu}$ includes not only independent triads but rather gives the general number of combinations of how $\nu's$ neighbors can be directly connected to each other.
Just as the average degree, the clustering coefficient changes its value depending on $\mathcal{M}_{i}$. For the ensembles of $\mathcal{T}_{i}$ as they are described above (in Sec.$\,$\ref{subsec:TRG}), some of the simple relations for the clustering coefficients are:
\begin{equation}
C_{\nu, 1} = C_{\nu, 2} = \frac{1}{2} C_{\nu, 7}, \qquad
C_{\nu, 3} = C_{\nu, 4} = C_{\nu, 5}
\end{equation}
The mean values for the clustering coefficient averaged over the whole network $\left\langle C\right\rangle$ are shown in Tab.$\,$\ref{tab:properties}. It is possible to tune the clustering coefficient within a certain range without changing the total number of triads, e.g. by gradually changing the network from $\mathcal{T}_{1}$ to $\mathcal{T}_{7}$. This can be done by replacing one $\mathcal{M}_{1}$ by one $\mathcal{M}_{7}$, thus tuning the fraction $\phi(T_1,T_7) = \frac{T_{1}-T_{7}}{T}$, where $T_{i}$ denotes the number of $\mathcal{M}_{i}$ triads in the network (unless stated otherwise, for all $\mathcal{T}_{i}$ we have $T_{i} = T =60$). 

\subsection{Paths and Centrality}
\label{subsec:diameter}
Paths and distances on networks are important in the context of disease dynamics. A path between nodes $\nu_1$ and $\nu_2$ represents the sequence of nodes that are passed by while following edges between $\nu_1$ and $\nu_2$. When individuals are densely connected the disease can spread quickly because of the short paths between the susceptible and infected. In particular, there are two kinds of paths that are notable: the shortest path (also known as \textit{distance}) and the longest path (also known as \textit{diameter}). The former considers a path between one pair of nodes and can be averaged over all pairs to find the mean distance, while the latter gives the maximum longest path between any pair of nodes in the network and is thus a global property. When two nodes are not connected by a path, the length of the path is formally set to $\infty$. Therefore, the diameter of a network which has more than one connected component is formally infinite, as well as the distance between two nodes of different connected components. At this point, it is worth mentioning that all TRGs used in this article have exactly one connected component.
Moreover, on directed graphs the distance between two nodes can still be infinite if there is no path along which all edges are directed. In the numerical example of TRGs with $T=60$, within $\mathcal{T}_{1}$ there are $30.2\,\pm\,4.1$ pairs with infinite distances, in $\mathcal{T}_{4}$: $47.2\,\pm\,4.1$ and in $\mathcal{T}_{5}$: $6.4\,\pm\,2.8$. The mean distances for different types of TRGs are given in Tab.$\,$\ref{tab:properties}.
Furthermore, we can provide some relations for the average diameter $\mathcal{D}_{i}$ of TRGs which differ only in the type of $\mathcal{M}_{i}$:
\begin{equation}
\begin{aligned}
\mathcal{D}_{1} \rightarrow \infty,\, \mathcal{D}_{4} &\rightarrow \infty,\, \mathcal{D}_{5} \rightarrow \infty \\
\mathcal{D}_{2} > \mathcal{D}_{3} &> \mathcal{D}_{6} > \mathcal{D}_{7},
\end{aligned}
\end{equation}
where the numbers in the subscripts denote again the type of the overrepresented triadic motifs. Note that for $\mathcal{T}_{1}$, $\mathcal{T}_{4}$ and $\mathcal{T}_{5}$ the diameter diverges, while for other $\mathcal{M}_{i}$ the diameter has finite values. The reason for this is associated with reciprocity and directionality of the TRGs.
Finally, a relevant measure for the significance of a node $\nu$ within a network is the betweenness centrality $B_{\nu}$ \cite{Danon2011}.
If $\nu$ is infected and its $B_{\nu}$ is high, the number of potential infection events during the next iteration step is higher than with a low $B_{\nu}$.

\subsection{Directionality}
\label{subsec:directionality}
TRGs of type $\mathcal{T}_{7}$ have only bidirectional edges because they are constructed only from subpatterns of type $\mathcal{M}_{7}$. Thus, they can also be considered as undirected. Most publications in the field of complex networks focus on undirected networks \cite{Satorras2015}. However, the presence of directed edges can influence significantly the dynamics of a system. For instance, the epidemic threshold was found to increase with an increasing fraction of directed to undirected edges on scale-free and binomial networks in SIS NIMFA dynamics \cite{epThreshold, epThreshold2}. This threshold is equal to $\frac{1}{\lambda}$, where $\lambda$ is the spectral gap, i.e. the largest real eigenvalue of the adjacency matrix $\boldsymbol{A}$. Furthermore, the network's assortativity $\rho$ \cite{Newman2002,Newman2003} may have an impact on the spectral gap \cite{epThreshold}. The values of $\lambda$ and $\rho$ for different $\mathcal{T}_{i}$ are shown in Tab.$\,$\ref{tab:properties}.\\
Moreover, an important global property of TRGs which comes with directionality is the occurrence of a hierarchical structure and, consequently, an inherent directionality. 
The latter, a term introduced by V. Dom\'inguez-Garc\'ia et al. \cite{Inherent}, states that the nodes of a network with a high inherent directionality can be sorted by a hierarchical order so that the network displays an average global direction towards which the nodes preferentially point. This hierarchical ordering can be computed using the equation
\begin{equation}
\label{eq:hierarchy}
h_{\nu} = 1 + \frac{1}{k_{in, \nu}} \Sigma_{j} A_{j\nu} h_{j},
\end{equation}
where $h_{\nu}$ ($k_{in,\nu}$) is the hierarchical level (in-degree), respectively, of the node $\nu$ to which node $j$ points. As a boundary condition, $h$ is set to zero for nodes with the lowest value of $k_{in}$, particularly for those with $k_{in} = 0$. Note that in Eq.$\,$\ref{eq:hierarchy} it is not always $h_{j} < h_{\nu}$, so that in order to examine whether a network shows a hierarchical structure, we calculate the \textit{current parameter} $\xi$ which gives the fraction of edges which are directed from nodes with lower towards nodes with higher $h$ \cite{Inherent} (see Tab.$\,$\ref{tab:properties}). Measurements of the current parameter $\xi$ have been performed on over 200 instances of every $\mathcal{T}_{i}$ with $T=60$. The highest value of $\xi$ is found for $\mathcal{T}_{1}$. The building blocks of this TRG are also known as \textit{feed-forward-loops} (FFLs). Networks with a high abundance of this kind of motifs are likely to possess a hierarchical structure and high inherent directionality \cite{Inherent}. On the contrast, TRGs with only reciprocal links, $\mathcal{T}_{7}$ or undirected triadic graphs, naturally display no preferential direction.
\\
\\
\\
\\
\section{Epidemics on preconstructed Triadic Random Graphs}
\label{sec:MeanField}
\subsection{Mean-field approaches}
The mathematical foundation of the dynamical process of disease transmission, the \textit{Susceptible-Infected-Recovered} model (SIR), was introduced by W. O. Kermack and A. G. McKendrick \cite{Kermack1927}. Since then, many optimizations were suggested \cite{Satorras2015}. Here, we will focus on the individual-based mean-field approach, which offers a simple way to include the network's topology through the adjacency matrix $\boldsymbol{A}$ \cite{Keeling2005}. Particularly, we will use the N-intertwined mean field approximation (NIMFA) introduced by P.V. Mieghem \cite{Mieghem2012,Mieghem2014} for the \textit{Susceptible-Infected-Susceptible} (SIS) model. It is given by a set of ordinary differential equations (ODEs) for the time evolution of the probability $x_{\nu}$ that $\nu$ is infected:
\begin{equation}
\frac{dx_\nu}{dt} = \beta (1-x_\nu) \Sigma_j A_{j\nu} x_j - \gamma x_\nu.
\label{eq:nimfa}
\end{equation}
where $\beta$ is the infection rate and $\gamma$ the recovery rate. A susceptible individual $\nu$ can be infected by other (infectious) individuals $j$ with the probability $\beta dt$ and subsequently recover (become again susceptible) with the probability $\gamma dt$. Both processes are assumed to be Poissonian. Eq.$\,$\ref{eq:nimfa} can be derived from the Markov theory, using the approximation that the total infection rate of a node is equal to the sum over all probable infection events from its neighbors \cite{Mieghem2011}. Although this model does not include a number of realistic assumptions (e.g. a discrete time-intervals for reinfections \cite{Gomez2010}), Eq.$\,$\ref{eq:nimfa} is computationally simple and it can be straightforwardly used with directed networks, such as Triadic Random Graphs. 
The fraction of infected individuals $x$ in the population of size 
$N$ can be calculated using the simple assumption that the total number of infected is $n_I = N \Sigma_{\nu}^N x_{\nu}$ and hence $x = \frac{n_I}{N} = \Sigma_{\nu}^N x_{\nu}$.\\
There are more models for disease transmission that include more realistic assumptions, e.g. that the number of individuals may change due to birth or death, or including further states such as being immune or vaccinated \cite{DiseaseDynamics}. However, the mathematical complexity increases with more realistic assumptions. In contrast, the SIS-model is rather simple but despite its simplicity it is still being investigated. For instance, an improvement to Eq.$\,$\ref{eq:nimfa} is to consider moment closure approximations of second order. This can be done by considering ODEs for pairs on infected nodes \cite{MieghemSecond}. However, this approximation is only valid for sufficiently large networks which is not the case for $N=49$. Hence, in this section we will mainly focus on the dynamics given by the first-order NIMFA (Eq.$\,$\ref{eq:nimfa}) and subsequently, in Sec.$\,$\ref{sec:stoch}, we will derive a novel stochastic approach which includes closure of higher order.

\subsection{Simulation results}
\begin{figure}
\includegraphics[width=1.0\linewidth]{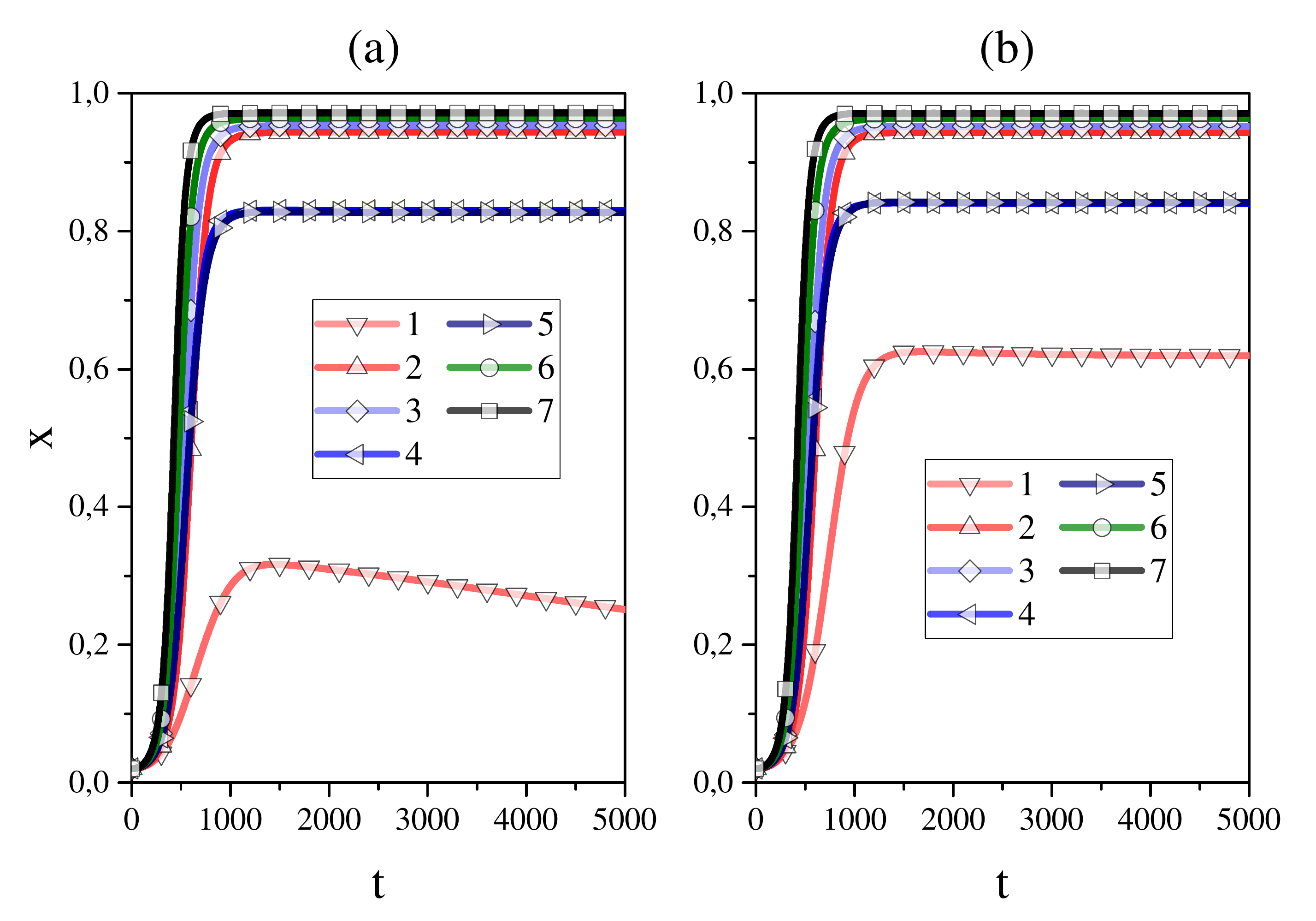}       
\caption[]{The change of the fraction of infected individuals $x$ with the number of iteration steps $t$ on (a) different non-randomized TRGs, (b) according null-models. The numbers in the legends correspond to the type of the motif $\mathcal{M}_{i}$ (see Fig.$\,$\ref{fig:motifs}) of which the TRGs consist. Note that the networks differ in nothing but the type of $\mathcal{M}_{i}$. The curves of $\mathcal{T}_{2}$, $\mathcal{T}_{3}$, $\mathcal{T}_{6}$, $\mathcal{T}_{7}$ coincide nearly perfectly with their null-models. The randomizations of $\mathcal{T}_{4}$ and $\mathcal{T}_{5}$ have a slightly higher $x_{end}$ than their non-randomized counterparts. The simulation results differ clearly between the original and the randomized versions of $\mathcal{T}_{1}$. Besides the value of $x$ at $t=5000$, the endemic state sets in for the null-model, on the contrast to the non-randomized $\mathcal{T}_{1}$, where $x$ decreases after reaching a maximum at $t\approx1300$. Parameters: $N=49$, $T=60$, $\beta=6$, $\gamma=1$, $x(0) = 0.02$.}   
\label{fig:nimfa60}    
\end{figure}

The set of ODEs in Eq.$\,$\ref{eq:nimfa} describes how the fraction of infected individuals evolves in time within a population of stationary size. Clustering and triadic substructures can have an impact on their time evolution. Therefore, we performed 4th-order-Runge-Kutta simulations on an ensemble of constructed networks ($\mathcal{T}_{i}$) which consist purely of distinctive closed triadic motifs $\mathcal{M}_{i}$. For every $\mathcal{T}_{i}$ we generated over 200 instances, each with $N=49$ and $T=60$, over which the simulation results were averaged.
Initially, the fraction of infected and susceptible individuals is $x(0) = \frac{1}{49}$ and $s(0) = 1 - x(0)$, respectively.
Fig.$\,$\ref{fig:nimfa60} shows the simulation outcome for the infection rate $\beta = 6$ and the recovery rate $\gamma = 1$.
\begin{figure}
\includegraphics[width=1.0\linewidth]{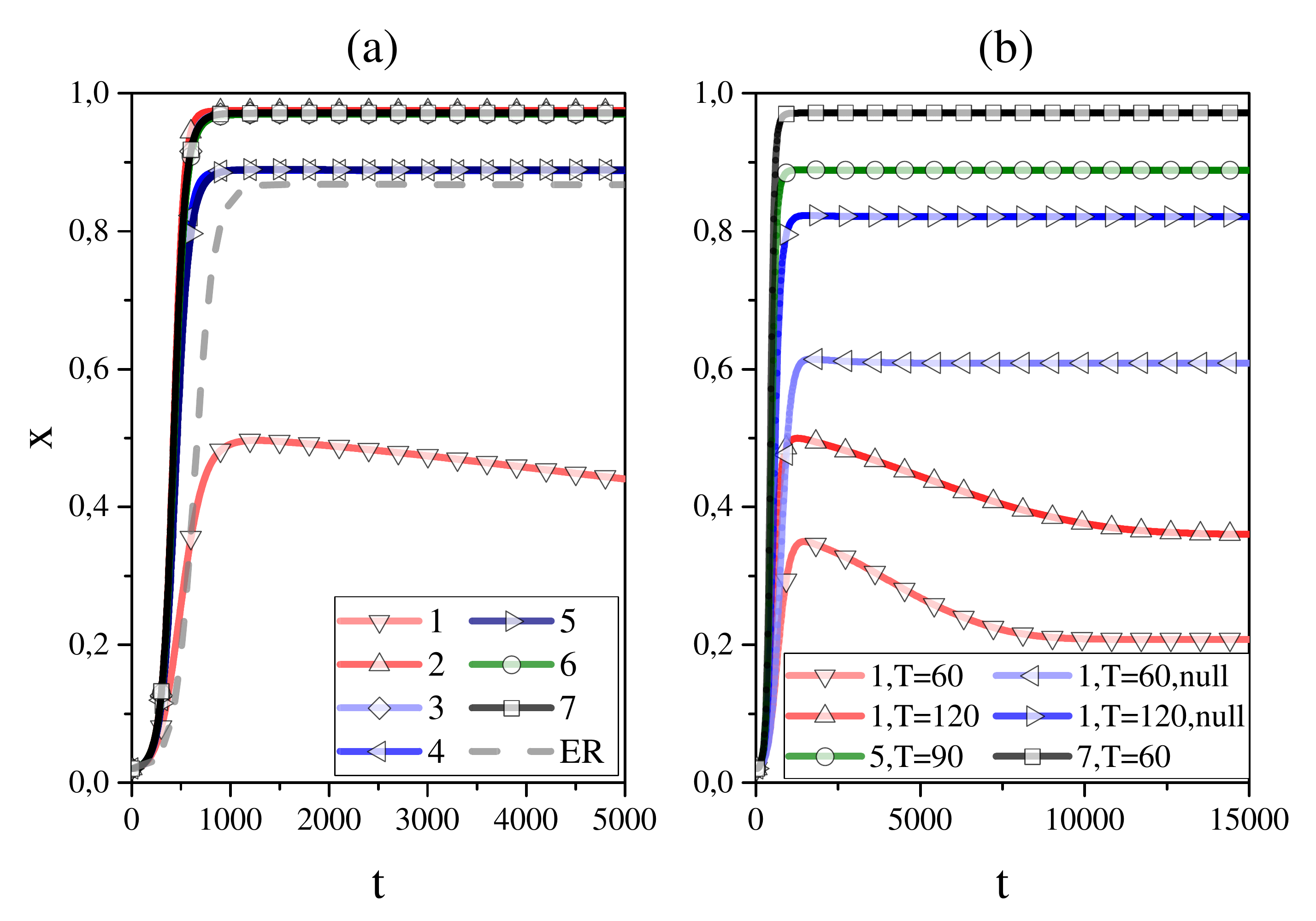}       
\caption[]{Mean-field simulation results for the change of the fraction of infected individuals $x$ with the number of iteration steps $t$ on (a) non-randomized $\mathcal{T}_{i}$ with different total numbers of triads $T_{i}$ and an Erd\H{o}s-R\'enyi graph with $E = 3T_{7}$, (b) different realizations of $\mathcal{T}_{1}$, $\mathcal{T}_{5}$ and $\mathcal{T}_{7}$. All TRGs in (a) have the same total number of edges $E$. The curves of $\mathcal{T}_{2}$, $\mathcal{T}_{3}$, $\mathcal{T}_{6}$, $\mathcal{T}_{7}$ coincide nearly perfectly, same holds for $\mathcal{T}_{4}$ and $\mathcal{T}_{5}$. The difference between the upper curves and the curves of $\mathcal{T}_{4}$ and $\mathcal{T}_{5}$ is approximately the fraction of nodes with $k_{in} = 0$ or $k_{out} = 0$. The simulation result for $\mathcal{T}_{1}$ is still clearly separated from the rest and cannot be as simply explained. In (b) results for $\mathcal{T}_{1}$ with $T=60$ and $T=120$ as well as corresponding null-models are shown on a longer timescale than in (a), curves of $\mathcal{T}_{5}$ ($\mathcal{T}_{7}$) with $T=90$ ($T=60$) are shown for comparison. The endemic state for $\mathcal{T}_{1}$ is reached at $t_{end}(T=60) \approx 9000$ and $t_{end}(T=120) \approx 13000$.  Parameters: $N=49$, $\beta=6$, $\gamma=1$, $x(0) = 0.02$.}   
\label{fig:regular}    
\end{figure} 

Except for the curves of $\mathcal{T}_{1}$, $\mathcal{T}_{4}$ and $\mathcal{T}_{5}$, all curves reach the endemic state (also known as metastable state, i.e. the non-trivial state where $x = x_{end}$ is constant, before the epidemic starts dying out) between $x_{end}=0.9$ and $x_{end}=1.0$. It is remarkable that even though the network is rather small and the triadic motifs clearly differ in their local structure as well as in their influence on the global properties of the networks, as presented in Sec.$\,$\ref{sec:TRG}, for most $\mathcal{T}_{i}$ their differences are barely reflected in the simulation outcome. On the other hand, $\mathcal{T}_{1}$, $\mathcal{T}_{4}$ and $\mathcal{T}_{5}$ stand out with $x_{end} \approx 0.85$ and $x(5000) \approx 0.25$, respectively. Furthermore, $\mathcal{T}_{1}$ displays a distinctive global maximum before it reaches the endemic state $x_{end} \approx 0.21$ (see Fig.$\,$\ref{fig:regular}). Consequently, a quite noteworthy conclusion is that replacing only one of the edges of the triadic motif $\mathcal{M}_{1}$ by a bidirectional edge significantly changes the outcome of an epidemic, while this effect is not as severe for other types of $\mathcal{M}_{i}$. All these observations hold qualitatively also for networks with lower values of $T$. For higher $T$ the dynamics approach the behavior of $\mathcal{T}_{7}$.\\
As one can see in Fig.$\,$\ref{fig:nimfa60}, the diseases spread the fastest (slowest) for $\mathcal{T}_{7}$ ($\mathcal{T}_{1}$). As discussed in Sec.$\,$\ref{subsec:directionality}, $\mathcal{T}_{7}$ can be considered as an undirected network. Hence, a susceptible node $\nu_S$ can be infected by any of its infectious neighbors and vice versa. The simulation results in Fig.$\,$\ref{fig:nimfa60} show that, on average, undirected graphs are a good approximation for most TRGs, except $\mathcal{T}_{1}$, $\mathcal{T}_{4}$ and $\mathcal{T}_{5}$. As discussed in the previous section, there are clear differences in the number of edges $E = T\cdot E(\mathcal{M}_{i})$ between $\mathcal{T}_{i}$. By changing the number of triads $T$ we can adjust $E$ according to Eq.$\,$\ref{eq:edges}. In Fig.$\,$\ref{fig:regular} it is shown that while the curves of $\mathcal{T}_{2}$, $\mathcal{T}_{3}$, $\mathcal{T}_{6}$ and $\mathcal{T}_{7}$ coincide up to small deviations, they are still clearly different to the curves of $\mathcal{T}_{4}$, $\mathcal{T}_{5}$ and $\mathcal{T}_{1}$.\\
Among all TRGs, $\mathcal{T}_{7}$ has the lowest diameter and distance, diseases can spread quickly through the network. Note also that we use only networks with exactly one connected component and the diameter of $\mathcal{T}_{7}$ has a finite value, thus all nodes can be reached by an infection. This is not necessarily the case for $\mathcal{T}_{1}$, $\mathcal{T}_{4}$ and $\mathcal{T}_{5}$. In fact, for all instances of $\mathcal{T}_{1}$ a divergent diameter and several pairs of nodes with a divergent distance are found. Consequently, there are nodes which can never be infected (unless initially), i.e. their in-degree is $k_{in}= 0$ or can never transmit an infection (see also Sec.$\,$\ref{subsec:degreeDistribution}). Nodes with $k_{in} = 0$ directly reduce $x$; nodes with $k_{out} = 0$ can only absorb but not share an infection. Thus, both kinds of nodes do not actively participate in the dynamics of the system and their presence reduces the density of infected individuals $x$.\\
Lastly, for $\mathcal{T}_{1}$, the far lower value of $x_{end}$ and the occurrence of a global maximum in $x$ is due to the occurrence of hierarchical structure, which we conclude from comparing the simulation outcome to randomized $\mathcal{T}_{1}$ in the next section.

\subsection{Comparison with Null-Models}
Remarkably, while considering the relations of the degree distribution or the clustering coefficient between different types of $\mathcal{T}_{i}$, we can exclude both properties from contributing significantly to the difference between the simulation outcomes. This becomes most evident when comparing the curves of $\mathcal{T}_{1}$ and $\mathcal{T}_{2}$. In order to examine whether other measures might have a greater impact, we compare the results to their corresponding null-models.\\
The lines in Fig.$\,$\ref{fig:regular} (b) show the simulated curves for randomized TRGs. For every instance of $\mathcal{T}_{i}$, 10 randomizations have been generated using the switching algorithm mentioned above (Sec.$\,$\ref{subsec:TRG}). Thus, the simulations are averaged over more than 2000 randomizations per $\mathcal{T}_{i}$. Except for $\mathcal{T}_{1}$, $\mathcal{T}_{4}$ and $\mathcal{T}_{5}$ all simulated curves of the null-models coincide perfectly with their non-randomized equivalents. Since the in- and out- degrees of every node are preserved, the number of nodes which cannot be infected or cannot transmit an infection does not change either. Hence, the curves of $\mathcal{T}_{1}$, $\mathcal{T}_{4}$ and $\mathcal{T}_{5}$ still have a significantly lower $x_{end}$ as compared to other TRGs. However, particularly the dynamics on $\mathcal{T}_{1}$ displays a clear difference to its randomized counterpart. This remarkable result shows that the local structure of the motifs can in fact have a significant impact on the dynamics and thus influence the epidemic spread without changing $N$,$E$ and the distribution of the degree or the clustering coefficient.\\
We assume that it is the presence of a hierarchical structure, which is distinctive for networks with an abundance of FFL-motifs \cite{Inherent}, that is responsible for the clearly different behavior of epidemic spread on $\mathcal{T}_{1}$. As we will see, this becomes more evident when comparing other measures of $\mathcal{T}_{1}$ shown in Tab.$\,$\ref{tab:properties} with those of other TRGs, particularly with $\mathcal{T}_{2}$ and $\mathcal{T}_{7}$.

\subsection{Fraction $\phi$}
\label{subsec:fraction}
In order to better understand why $x$ evolves on $\mathcal{T}_{1}$ so differently from other $\mathcal{T}_{i}$, it is worthwhile to examine how the fraction of infected individuals changes when a TRG approaches the form of $\mathcal{T}_{1}$. Particularly, we are interested in the transition from a purely undirected graph to a graph with an abundance of FFL-motifs, which can be simulated by changing the fraction $\phi(T_1,T_7) = \frac{T_{1}-T_{7}}{60}$, i.e. the numbers of $\mathcal{M}_{1}$ and $\mathcal{M}_{7}$ triads within the network. We generated TRGs where, starting with 60 $\mathcal{M}_{7}$-motifs (i.e. $T_{7} = 60$) out of $T=60$ triads, i.e. $\phi(0,60)=-1.0$, we replace one $\mathcal{M}_{7}$-motif by one $\mathcal{M}_{1}$-motif and subsequently perform the same simulations as in Fig.$\,$\ref{fig:nimfa60} (again, for every value of $\phi(T_i,T_j)$ we generate an ensemble of TRGs with $>$200 samples).
\begin{figure}
 \includegraphics[width=1.0\linewidth]{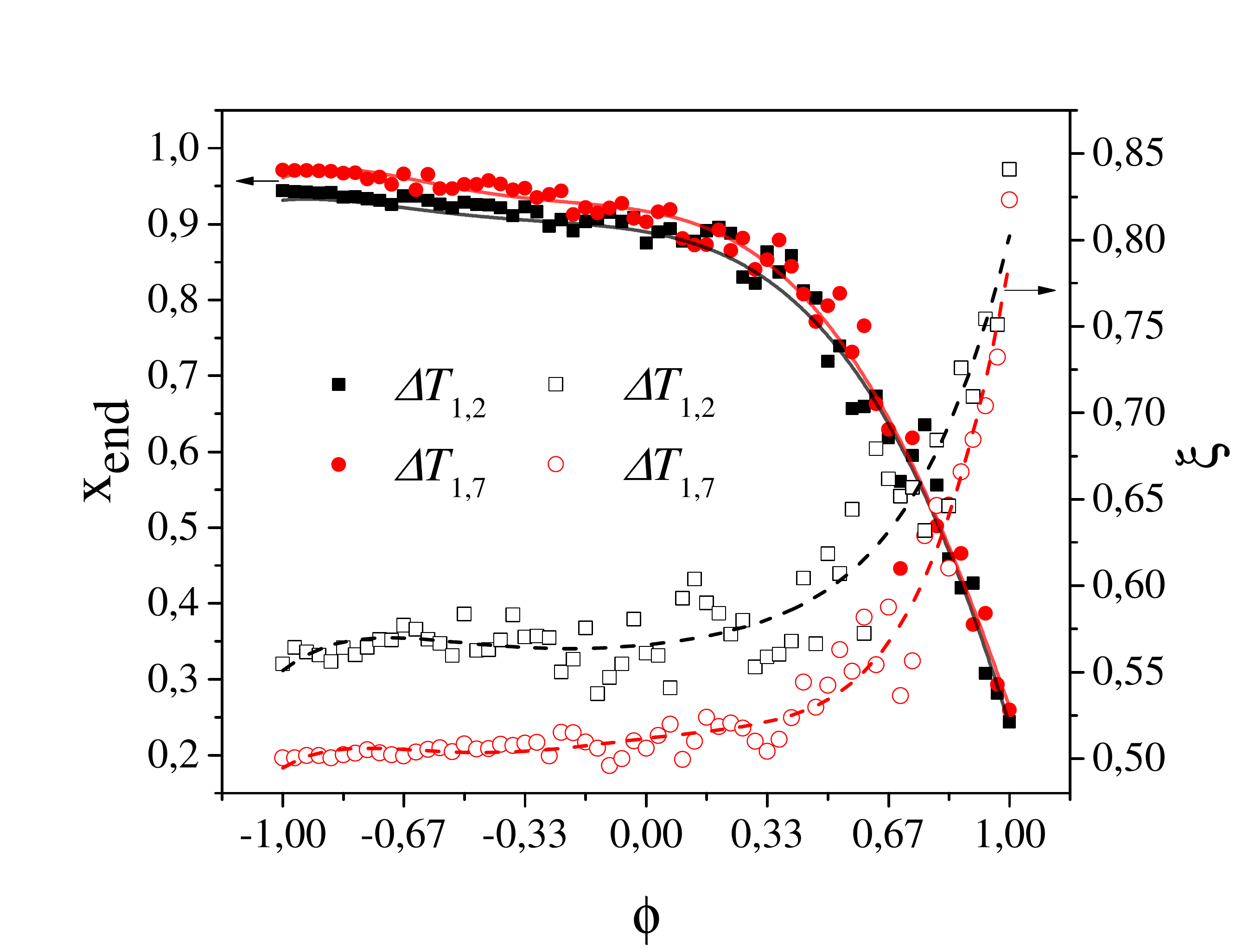}
 \caption{\label{fig:fraction} Bold symbols (\textbf{left} y-axis): Change of $x_{end}$ with a decreasing number of $\mathcal{M}_{2}$-motifs (squares) and $\mathcal{M}_{7}$-motifs (circles). Empty symbols (\textbf{right} y-axis): Increasing hierarchical structure, represented by the current parameter $\xi$ with a decreasing number of $\mathcal{M}_{2}$-motifs (squares) and $\mathcal{M}_{7}$-motifs (circles). $\Delta T_{i,j} = T_{i} - T_{j}$. All lines are guides to the eye.}
\end{figure}
The values of $x_{end}$ in the endemic state for different $\phi(T_1,T_7) = \frac{T_{1}-T_{7}}{60}$ and $\phi(T_1,T_2) = \frac{T_{1}-T_{2}}{60}$ are shown in Fig.$\,$\ref{fig:fraction}. The change of $x_{end}$ for a transition from a graph with an abundance of $\mathcal{M}_{2}$ (this motif is also known as \textit{feed-back-loop}) to $\mathcal{T}_{1}$ is very similar to the transition from $\mathcal{T}_{7}$. In contrast to the latter, $\mathcal{T}_{2}$ has only unidirectional links and the total number of edges is $E_{2} = E_{1}$. The fraction of infected individuals in the endemic state decreases for an increasing number of $\mathcal{M}_{1}$ in both cases.\\
The only measures of $\mathcal{T}_{1}$ which differ clearly from all other $\mathcal{T}_{i}$ are: betweenness centrality $\left\langle B\right\rangle$, graph spectrum $\lambda$, assortativity $\rho$ and current parameter $\xi$ (see Tab.$\,$\ref{tab:properties}). However, $\left\langle B\right\rangle_{1}$ has a relatively high deviation from the expectation value and a high variety of values for every network and even though $\Delta \left\langle B\right\rangle_{2,7} = \left\langle B\right\rangle_{2} - \left\langle B\right\rangle_{7} > \Delta \left\langle B\right\rangle_{2,1} > \Delta \left\langle B\right\rangle_{7,1}$ it is not reflected in the values of $x_{end}$ (subscript corresponds to the type of $\mathcal{T}_{i}$). Similar arguments suggest that the change of $x_{end}$ cannot be related to $\lambda$ and $\rho$. On the contrary, the current parameter $\xi_{i}$ is close to 0.5 for all $\mathcal{T}_{i}$, except $\mathcal{T}_{5}$ and $\mathcal{T}_{1}$. As mentioned before (Sec.$\,$\ref{subsec:directionality}), when $\xi \approx 0.5$ the network has nearly no hierarchical ordering but when $\xi \approx 1.0$ the network structure is purely hierarchical. Fig.$\,$\ref{fig:fraction} shows the nonlinear increase of $\xi$ for an increasing $\phi(T_1,T_7)$ or $\phi(T_1,T_2)$. Moreover, for the null-model of $\mathcal{T}_{1}$, the value of the current parameter is $\xi_{1, null} = 0.711 \pm 0.068$ and thus lower than for the non-randomized $\mathcal{T}_{1}$. Therefore, from the analysis of epidemic dynamics on TRGs we can conclude that within the SIS-model the set in of hierarchical structure, together with the presence of nodes with $k_{in} = 0$ or $k_{out} = 0$, leads to lower fraction of infected individuals.

\section{Node-tagging approach}
\label{sec:stoch}
\subsection{CTMC for regular graphs and Erd\H{o}s-R\'enyi-Graphs}
\label{subsec:regGrERG}
Besides the results of the previous sections, in order to examine whether the local structure of triadic motifs influences the epidemic dynamics, it may be also useful to consider an approach which does not include whole preconstructed networks in form of an adjacency matrix. Instead, an investigation from the perspective of a model which focuses only on the type and number of triadic motifs may be advantageous. A realization of such a model is presented in the next subsection (Sec.$\,$\ref{sec:statTRG}). It is based on the Continuous Time Markov Chain (CTMC) and Kolmogorov equations. TRGs which are discussed in this work are relatively small systems, so that neither the number of nodes nor the number of triads can be considered continuous. Furthermore, the spread of diseases throughout all our simulations takes place on static networks, i.e. the topology of our networks does not change with time. This is a good approximation for infections which spread much faster than the change of the network-topology \cite{Danon2011}. Thus, when looking at the dynamics of the epidemic we can assume very small time steps. Finally, the epidemic spread is assumed to be a Markov process, i.e. the system has no memory and the dynamics of time $t+dt$ is not influenced by the dynamics at $t$. With these assumptions (continuous spread of diseases, discrete variables (nodes, triads), Markov assumption) the choice of CTMC is a suitable stochastic model \cite{Keeling2008}. When the variables can be assumed to be continuous, it is more appropriate to use stochastic differential equations. As a simple example for the analysis of epidemic dynamics we will first discuss the SIS-model on a regular graph, i.e. a graph where all nodes have the same total degree $k_{tot}$. A simple deterministic formulation of the change of infected individuals is given by 
\begin{equation}
\label{MeanFieldI}
\frac{dn_I}{dt} = \beta n_S n_I - \gamma n_I,
\end{equation}
where $\beta$ is the infection rate and $\gamma$ the recovery rate.
Let us denote $p_{n_I}$ the probability of finding $n_I$ infected individuals at time $t$. The number of the susceptible individuals is then given by $n_S = N - n_I$ ($N$ is the total number of individuals). Within the framework of the CTMC model, $p_{n_I}$ changes with time according to the Kolmogorov forward equations \cite{Keeling2008} 
\begin{equation}
\label{KolmogorovSimple}
\frac{dp_{n_I}}{dt} = p_{n_I-1}w_{n_I-1}^++p_{n_I+1}w_{n_I+1}^--p_{n_I}\left(w_{n_I}^++w_{n_I}^-\right),
\end{equation}
where we used the transition rates $w_{n_I}^+ = \beta n_Sn_I$ and $w_{n_I}^- = \gamma n_I$ for the forward and backward transtitions, respectively. \\
The formulation of equations (\ref{MeanFieldI}) and (\ref{KolmogorovSimple}) uses the mean-field approximation that the expectation value of the \textit{susceptible-infected pairs} is equal to their product, i.e. $\langle n_{SI}\rangle = \langle n_S\rangle\langle n_I\rangle / N$.
This rather ideal assumption can be refined by the \textit{pairwise closure} by deriving differential equations for pairs of individuals, that is for $n_{SS}$, $n_{SI}$ and $n_{II}$ \cite{NewmanBook, House2009, Eames2002}. Furthermore, stochastic ordinary differential equations for the pairwise-based approach have been presented \cite{Dangerfield2008}, taking into account that a change of the state of a node (i.e. becoming infectious or susceptible) quantitatively influences $n_{SS}$, $n_{SI}$ and/or $n_{II}$. To elaborate the latter point, let us consider a susceptible node $\nu_{S}|_{ t_1}$ which has $k_S$ susceptible and $k_I$ infected neighbors at time $t_1$. Thus, $\nu_S$ and its nearest surrounding make up for $k_S$ $(SS)$-pairs and $k_I$ $(SI)$-pairs. Consequently, an infection of $\nu_S$ (i.e. the transition $\nu_{S}\vert_{ t_1} \rightarrow \nu_{I}\vert_{ t_2}$) results in the annihiliation of $k_S$ $(SS)$-pairs ($n_{SS}\vert_{ t_2} = n_{SS}\vert_{ t_1} - k_S$), creation of $k_I$ $(II)$-pairs ($n_{II}\vert_{ t_2} = n_{II}\vert_{ t_1} - k_I$) as well as annihiliation of $k_S$ and creation of $k_I$ $(SI)$-pairs ($n_{SI}\vert_{ t_2} = n_{SI}\vert_{ t_1} + k_S - k_I$). In other notation we can write
\begin{equation}
\boldsymbol{v}_{st}\vert_{ t_2} = \boldsymbol{v}_{st}\vert_{ t_1} + \boldsymbol{v}_{sh}\vert_{ t_1},
\end{equation} 
where $\boldsymbol{v}_{st}\vert_{ t_1} = (n_S, n_I, n_{SS}, n_{SI}, n_{II})$ is the \textit{state vector} at time $t_1$ and $\boldsymbol{v}_{sh} = (-1, 1, -k_S, k_S - k_I, k_I)$ the \textit{shift vector} of the forward transition (notation is largely compatible with \cite{Noel2014}). For the backward transition $\boldsymbol{v}_{sh}$ changes its sign. Following this approach, it is important to calculate the probability of finding the central node with exactly $k_S$ susceptible and $k_I$ infected neighbors. In a regular network where every node has the same total degree $k_{tot}$ and for a susceptible central node $\nu_S$ this probability is given by \cite{Dangerfield2008}
\begin{equation}
p_{\nu_S}\left(\boldsymbol{v}_{st}, \boldsymbol{v}_{sh}\right) = \binom{k_{tot}}{k_I} \left(\frac{n_{SI}}{k_{tot}n_S}\right)^{k_I}\left(1-\frac{n_{SI}}{k_{tot}n_S}\right)^{k_S}.
\end{equation}
Similarly, the probability for an infected node $\nu_I$ is given by 
\begin{equation}
p_{\nu_I}\left(\boldsymbol{v}_{st}, \boldsymbol{v}_{sh}\right) = \binom{k_{tot}}{k_I} \left(\frac{n_{II}}{k_{tot}n_I}\right)^{k_I}\left(1-\frac{n_{II}}{k_{tot}n_I}\right)^{k_S}.
\end{equation}
Using these probabilities we can construct the transition rates as 
\begin{equation}
w^+\left(\boldsymbol{v}_{st}, \boldsymbol{v}_{sh}\right) = \beta p_{\nu_S}\left(\boldsymbol{v}_{st}, \boldsymbol{v}_{sh}\right)
\label{eq:transRateSI}
\end{equation}
\begin{equation}
w^-\left(\boldsymbol{v}_{st}, \boldsymbol{v}_{sh}\right) = \gamma\,p_{\nu_I}\left(\boldsymbol{v}_{st}, \boldsymbol{v}_{sh}\right).
\label{eq:transRateII}
\end{equation}
The change of pairs with time is then given by
\begin{equation}
\frac{dn_{SI}}{dt} = \Sigma_{k_I = 0}^{k_{tot}}\left(k_S-k_I\right) \left[w^+\left(\boldsymbol{v}_{st}, \boldsymbol{v}_{sh}\right) - w^-\left(\boldsymbol{v}_{st}, \boldsymbol{v}_{sh}\right)\right]
\label{eq:SODESI}
\end{equation}
\begin{equation}
\frac{dn_{II}}{dt} = \Sigma_{k_I = 0}^{k_{tot}}k_I \left[w^+\left(\boldsymbol{v}_{st}, \boldsymbol{v}_{sh}\right) - w^-\left(\boldsymbol{v}_{st}, \boldsymbol{v}_{sh}\right)\right]
\label{eq:SODEII}
\end{equation}
\begin{equation}
n_{SS} = N k_{tot} - n_{SI} + n_{II}.
\label{eq:SODESS}
\end{equation} \\
This approach can be easily extended from regular to ER graphs by adjusting the probabilities within the transition rates. In ER graphs the distribution of the node's degree is binomial. Hence, the probability of finding the susceptible central node $\nu_S$ surrounded by $k_S$ susceptible and $k_I$ infected neighbors can be obtained from the binomial distribution \cite{Noel2014}:
\begin{eqnarray}
&p_{\nu_S}\left(\boldsymbol{v}_{st}, \boldsymbol{v}_{sh}\right) = \binom{2n_{SS}}{k_S} \left(\frac{1}{n_S}\right)^{k_S}\left(1-\frac{1}{n_S}\right)^{2n_{SS}-k_S}  \nonumber \\*  
&\cdot \binom{n_{SI}-1}{k_I-1} \left(\frac{1}{n_S}\right)^{k_I-1}\left(1-\frac{1}{n_S}\right)^{n_{SI}-k_I}.
\end{eqnarray}
$p_{\nu_I}\left(\boldsymbol{v}_{st}, \boldsymbol{v}_{sh}\right)$ can be derived accordingly.\\
Next, we will make use of the fact that $\mathcal{T}_{i}$ are constructed purely from closed triadic motifs $\mathcal{M}_{i}$ and derive a novel stochastic approach for this type of graphs.

\subsection{Node-tagging approach for TRGs}
\label{sec:statTRG} 
\begin{figure}                                
\begin{center}        
\includegraphics[width=0.6\linewidth]{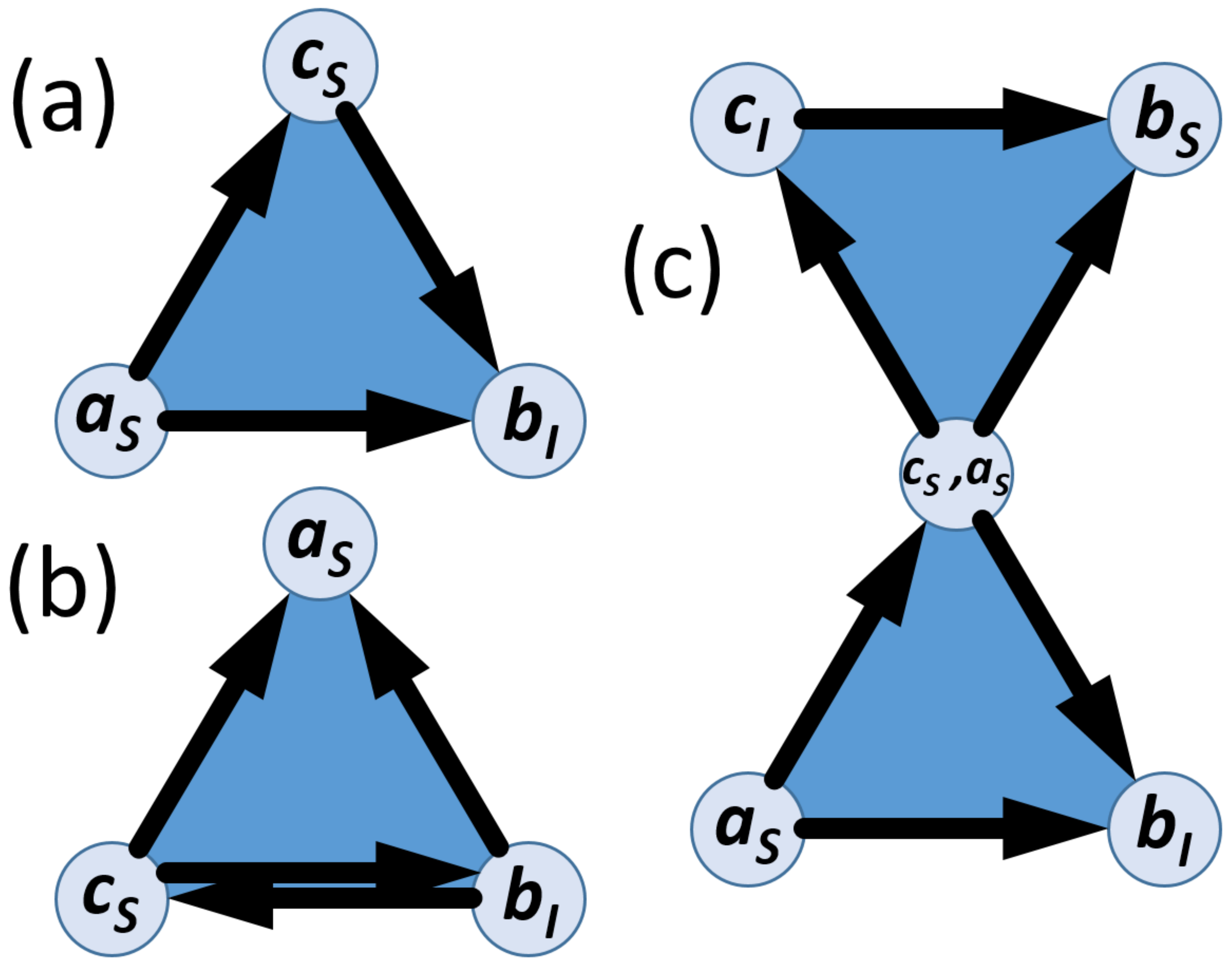}       
\caption[]{(a) Feed-Forward-Loop $\mathcal{M}_{1}$. (b) Triadic motif $\mathcal{M}_{4}$. Note that the edge between $c_S$ and $b_I$ is bidirectional, on the contrary to the unidirectional edge in (a). (c) Subgraph constructed from two triadic motifs of type $\mathcal{M}_{1}$ with the central susceptible node having a $c$-tag from the lower motif and an $a$-tag from the upper motif.}   
\label{M3M6}                                      
\end{center}
\end{figure}

\begin{figure}                                 
\begin{center}        
\includegraphics[width=0.8\linewidth]{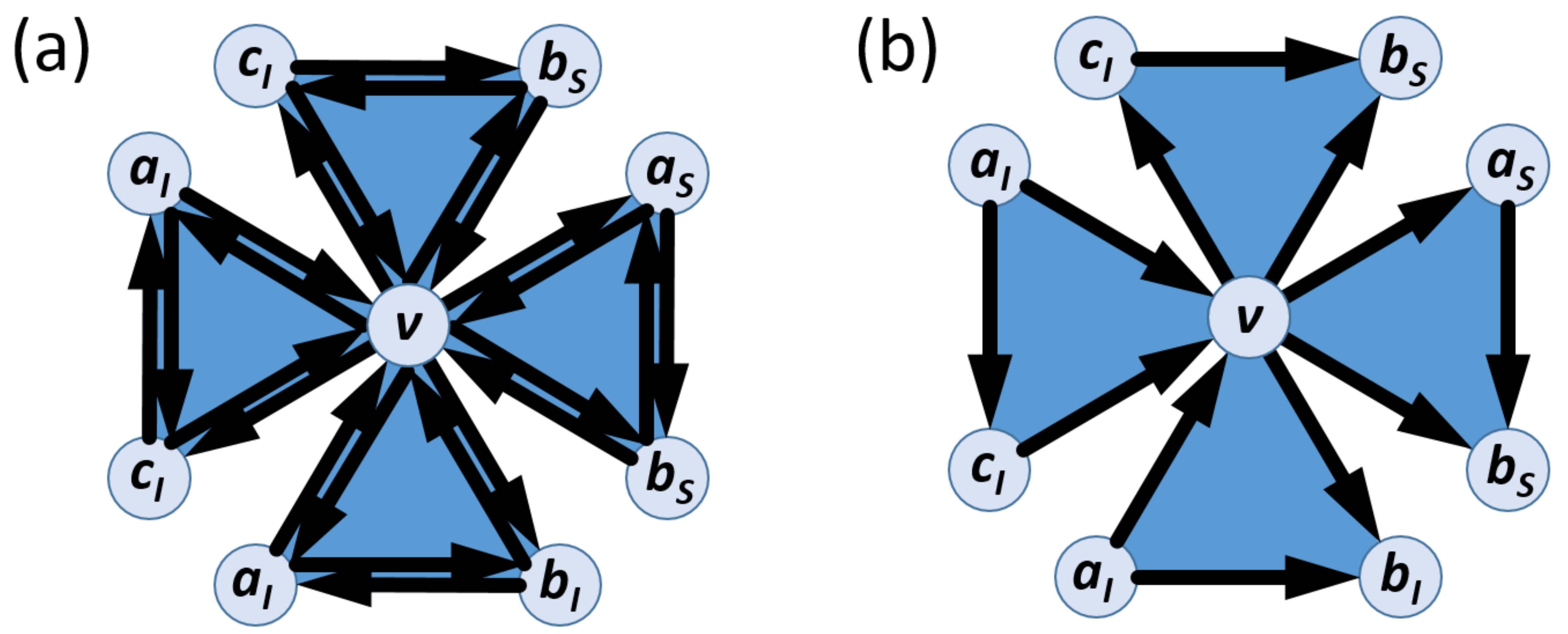}       
\caption[]{(a) High-order-motif with four $\mathcal{M}_{7}$-triads. (b) High-order-motif with four $\mathcal{M}_{1}$-triads. The probability for $\nu_S$ to get infected is higher in (a) then in (b).}   
\label{fig:M7M1}                                      
\end{center}
\end{figure}

In the above subsection, Sec.$\,$\ref{subsec:regGrERG}, we discussed existing stochastic models which aim to refine the stochastic analysis of epidemic spreading on simple networks by simplifying the consideration of possible effects of local clustering and heterogeneities \cite{Eames2002, Dangerfield2008, Noel2014}. There are more approaches in the literature which include stochastic models with pairwise closure as well as higher orders of closure \cite{Volz2011, House2009, Miller2014}. The approach used in the previous subsection works for large ($N \geq 10^3$) populations and is defined for undirected graphs. However, the networks considered in this article are small ($N = 49$) and directed. In order to consider stochastics of TRGs, we will extend the models discussed in the previous subsection by adjusting the transition rates $w^+\left(\boldsymbol{v}_{st}, \boldsymbol{v}_{sh}\right)$ and $w^-\left(\boldsymbol{v}_{st}, \boldsymbol{v}_{sh}\right)$ accordingly. For this purpose we will introduce a new formulation which is based on assigning characteristic \textit{tags} to nodes of each triadic motif $\mathcal{M}_{i}$.\\
Kashtan et al. \cite{Kashtan2004} defined in their work the \textit{role} of a node which describes this node's \textquotedblleft uniqueness \textquotedblright within a motif. For triads, two or three nodes have the same role if they can be permuted, while keeping their corresponding $(k_{in}, k_{out})$ and without changing the triad strucutre. One can see that there is a clear relation between the role of a node and its in- and out-degrees within a motif.
Consider the FFL $\mathcal{M}_{1}$ shown in Fig.$\,$\ref{M3M6} a). 
Each node in this triad has a different set of $\left(k_{in}, k_{out}\right)$. The node tagged as $\boldsymbol{a}$ corresponds to the tuple $(k_{in} = 0, k_{out} = 2)$, $\boldsymbol{b}$  to $(2, 0)$ and $\boldsymbol{c}$ to $(1, 1)$. When we add another $\mathcal{M}_{1}$ triad, any new nodes are tagged accordingly and the central node gets a second tag (see Fig.$\,$\ref{M3M6} (c)). Continuing this approach and extending it to triadic motifs of other types, we introduce three rules of the node-tagging (NoTa) approach:
\begin{enumerate}
\item Within each triad $\mathcal{M}_{i}$ there is always exactly one node tagged $a$, one $b$ and one $c$.
\item Each tag-index (a,b or c) corresponds to the same set $(k_{in},k_{out})$ in every triadic motif of the network.
\item The tags are additive for every node, i.e. for each node of a network there is a corresponding tag-triple $(\kappa_{a} a,\kappa_{b} b,\kappa_{c}c)$ with $\kappa_{a},\kappa_{b},\kappa_{c}\in\mathbb{N}_0$. 
\end{enumerate}
Using this formalism we can also assign tags to the nodes of the triad $\mathcal{M}_{4}$ even if two nodes have an equal set of $\left(k_{in}, k_{out}\right)$, as shown in Fig.$\,$\ref{M3M6} (b). This does not violate the rules because $(k_{in},k_{out})$ need not differ within a triad. Consequently, in the motif $\mathcal{M}_{7}$ all three tags correspond to the same set $(k_{in}=2,k_{out}=2)$. This approach allows to easily connect the local structure of a motif to the global structure of $\mathcal{T}_{i}$. The former can be described by a motif matrix $\boldsymbol{D}_{i}$ ($i$ corresponds to the index of $\mathcal{M}_{i}$), where 
$$ d_{\alpha\beta}=\left\{
\begin{array}{ll} 
	1, & \text{if } \alpha \text{ can transmit the infection to }\beta \\
    0, & \text{otherwise}
\end{array}\right.,$$
where $\alpha$ and $\beta$ represent tags.
Let's consider for example the triad $\mathcal{M}_{4}$. Without loss of generality we can assign tags to the nodes of this triad as in Fig.$\,$\ref{M3M6} (b), in which case $(k_{in},k_{out})_{a} \neq (k_{in},k_{out})_{b} = (k_{in},k_{out})_{c}$. The corresponding motif matrix is 
\begin{equation}
 \boldsymbol{D}_4 =
\begin{pmatrix}
0 & 0 & 0 \\
1 & 0 & 1 \\
1 & 1 & 0 
\end{pmatrix}.
\end{equation}
Using this notation we can define the structure of any other triad. The concept of a motif matrix is not new and has been introduced previously by Milo et al. \cite{Kashtan2004}. Note that $Tr(\boldsymbol{D})=0$ for each $\boldsymbol{D}_{i}$ because we exclude self-loops in all our networks. To complete the depiction of the triad, we assign a state to its nodes ($S$ for susceptible and $I$ for infected), so that $\mathcal{M}_{4} = (a_Sb_Ic_S)_4$ fully describes the triad shown in Fig.$\,$\ref{M3M6} (b). For simplicity, we will call a node in a susceptible state and with a $b$-tag: a $b_S$-tag (and similarly for other tags and infected states).\\
Having defined a notation for the triads we move on to considering the motifs of higher order than triadic motifs. Within the following stochastic model a \textit{high-order-motif} consists of a central node $\nu$ which is part of $\theta$ triads, i.e. it has $2\,\theta$ neighbors. 
As an approximation, we neglect all triads of $\nu$'s neighbors in which $\nu$ does not participate. The tags of $\nu$ can be now precisely determined by the tags of its neighbors. That is, if $\nu$ has exactly one neighbor having the $a$-tag and exactly one neighbor having the $b$-tag it immediately means that $\nu$ has exactly one tag and it is $c$. This way it suffices to know the tags of $\nu$'s neighbors which are given by the vector $\boldsymbol{\chi}$ in Eq.$\,$\ref{eq:chi}, with $dim(\chi)=12$. Furthermore, we are now able to describe our system with a new state vector $\boldsymbol{v}_{st}$ (see Eq.$\,$\ref{eq:vst}), with $dim(x)=8$. From $\boldsymbol{v}_{st}$ and $\boldsymbol{D}$ the number of $(SI)$-pairs, where $I$ points to $S$ can be calculated using Eq.$\,$\ref{eq:xSI}.
\begin{widetext}
\begin{eqnarray}
\boldsymbol{\chi} &=& (\chi_{a_Sb_S},\chi_{a_Sc_S},\chi_{b_Sc_S},\chi_{a_Ib_S},\chi_{a_Ic_S},\chi_{b_Ia_S},\chi_{b_Ic_S},\chi_{c_Ia_S},\chi_{c_Ib_S},\chi_{a_Ib_I},\chi_{a_Ic_I},\chi_{b_Ic_I})^{T}\label{eq:chi} \\
\boldsymbol{v}_{st} &=& (n_{A_SB_SC_S},n_{A_SB_SC_I}, n_{A_SB_IC_S}, n_{A_IB_SC_S}, n_{A_SB_IC_I}, n_{A_IB_SC_I}, n_{A_IB_IC_S}, n_{A_IB_IC_I})^{T}\label{eq:vst} \\
n_{SI} &=& \boldsymbol{v}_{st} \cdot \left(0,d_{ca}+d_{cb},d_{ba}+d_{bc},d_{ab}+d_{ac},d_{ca}+d_{ba},d_{ab}+d_{cb},d_{ac}+d_{bc},0\right)^{T}\label{eq:xSI}
\end{eqnarray}
\end{widetext}
In this notation, $n_{A_SB_IC_S}$ denotes the number of all triads in the network with susceptible $a$- and $c$-tags and an infected $b$-tag. Furthermore, $\chi_{a_Sb_I}$ denotes a triad in which $\nu$ has a neighbor with an $a_S$-tag and a neighbor with a $b_I$ tag. As we will see later, the vector $\boldsymbol{\chi}$ allows us to include into our stochastic analysis the contribution of both, directionality and local structure, i.e. the type of $\mathcal{T}_{i}$.\\
An infection of the susceptible central node $\nu$ leads to the reduction of $n_{A_SB_SC_S}$ by $\chi_{a_Sb_S},\chi_{a_Sc_S}$ and $\chi_{b_Sc_S}$ neighbors of $\nu$. Thus, the first component of the shift vector is $- (\chi_{a_Sb_S} + \chi_{a_Sc_S}+\chi_{b_Sc_S})$. At the same time $\chi_{a_Sb_S}$ new $(a_Sb_Sc_I)$-triads are created and $(\chi_{c_Ia_S} + \chi_{c_Ib_S})$ of $(a_Sb_Sc_I)$-triads are destroyed. The change of other components of the state vector can be derived in the same way. This procedure results in the change of the state which is given by the shift vector 
\begin{equation}
 \boldsymbol{v}_{sh} =
\begin{pmatrix}
-(\chi_{a_Sb_S}+\chi_{a_Sc_S}+\chi_{b_Sc_S}) \\
\chi_{a_Sb_S} - (\chi_{c_Ia_S}+\chi_{c_Ib_S}) \\
\chi_{a_Sc_S} - (\chi_{b_Ia_S}+\chi_{b_Ic_S}) \\
\chi_{b_Sc_S} - (\chi_{a_Ib_S}+\chi_{a_Ic_S}) \\  
\chi_{b_Ia_S}+\chi_{c_Ia_S} - \chi_{b_Ic_I} \\
\chi_{a_Ib_S}+\chi_{c_Ib_S} - \chi_{a_Ic_I} \\
\chi_{a_Ic_S}+\chi_{b_Ic_S} - \chi_{a_Ib_I} \\
\chi_{a_Ib_I}+\chi_{a_Ic_I}+\chi_{b_Ic_I} 
\end{pmatrix}
\end{equation}
Additionally, we can track the changes to the total numbers of the infected tags in the network $n_{A_I}$, $n_{B_I}$ and $n_{C_I}$ because they are given by the components of $\boldsymbol{v}_{st}$:
\begin{eqnarray}
n_{A_I} &=& n_{A_IB_SC_S} + n_{A_IB_IC_S} + n_{A_IB_SC_I} + n_{A_IB_IC_I} \nonumber \\*
n_{B_I} &=& n_{A_SB_IC_S} + n_{A_SB_IC_I} + n_{A_IB_IC_S} + n_{A_IB_IC_I} \\*
n_{C_I} &=& n_{A_SB_SC_I} + n_{A_SB_IC_I} + n_{A_IB_SC_I} + n_{A_IB_IC_I}. \nonumber
\label{eq:xAI}
\end{eqnarray}
Next, similarly to Sec.$\,$\ref{subsec:regGrERG}, we define the transition rates $w^+(\boldsymbol{v}_{st}, \boldsymbol{v}_{sh})$ and $w^-(\boldsymbol{v}_{st}, \boldsymbol{v}_{sh})$ for forward and backward transitions:
\begin{equation}
w^+(\boldsymbol{v}_{st}, \boldsymbol{v}_{sh}) = \beta\, p_{\nu_S}(\boldsymbol{v}_{st}, \boldsymbol{v}_{sh})
\label{transRateForwardRole}
\end{equation}
\begin{equation}
w^-(\boldsymbol{v}_{st}, \boldsymbol{v}_{sh}) = \gamma\, p_{\nu_I}(\boldsymbol{v}_{st}, \boldsymbol{v}_{sh}).
\label{transRateBackwardRole}
\end{equation} \\
The probabilities are given by the products of distributions
\begin{eqnarray}
p_{\nu_S}(\boldsymbol{v}_{st}, \boldsymbol{v}_{sh},\boldsymbol{\chi}) = \Pi_{\chi_i} p_{\nu_S,\chi_i}(\boldsymbol{v}_{st}, \boldsymbol{v}_{sh},\chi_i) \\
p_{\nu_I}(\boldsymbol{v}_{st}, \boldsymbol{v}_{sh},\boldsymbol{\chi}) = \Pi_{\chi_i} p_{\nu_I,\chi_i}(\boldsymbol{v}_{st}, \boldsymbol{v}_{sh},\chi_i).
\end{eqnarray}
Here, $p_{\nu_S,\chi_i}(\boldsymbol{v}_{st}, \boldsymbol{v}_{sh})$ gives the probability for $\nu_S$ of having $\chi_i$ pairs of neighbors of type $i$. As an example, let us consider the probability for the central node $\nu_S$ to have $\chi_{a_Sb_I}$ neighboring pairs of type $a_Sb_I$, i.e. that $\nu_S$ has $\chi_{a_Sb_I}$ times the tag $c_S$, $\chi_{a_Sb_I}$ infectious neighbors with the tag $b$ and $\chi_{a_Sb_I}$ susceptible neighbors with the tag $a$. This can be approximated by the probability that out of $n_{A_SB_IC_S}$-triads we pick the one with the susceptible central node $\nu_{c_S}$ exactly $\chi_{a_Sb_I}$-times, which is given by the binomial distribution
\begin{eqnarray}
&p_{\nu_S,\chi_{a_Sb_I}}(\boldsymbol{v}_{st}, \boldsymbol{v}_{sh},\chi_{a_Sb_I}) = \nonumber \\*
&\binom{n_{A_SB_IC_S}}{\chi_{a_Sb_I}} \left(\frac{1}{C_S}\right)^{\chi_{a_Sb_I}}\left(1-\frac{1}{C_S}\right)^{n_{A_SB_IC_S}-\chi_{a_Sb_I}}
\label{eq:pnuSAsBi}
\end{eqnarray}
The elements of $\boldsymbol{\chi}$ are chosen randomly from uniform distributions. However, as we consider a relatively small system ($N=49$, $T=60$), it is important to take into account that at every time step the elements of $\boldsymbol{\chi}$ have upper bounds which depend on $\boldsymbol{v}_{st}$ and the total number of $\nu$'s triads $\theta$. For instance, $\chi_{a_Sb_I}$ must not exceed the total number of triads which may include it, i.e. $\chi_{a_Sb_I}~\leq~min\left\lbrace\left(n_{A_SB_IC_S} + n_{A_SB_IC_I}\right),\theta\right\rbrace$.\\
Moreover, the rate in Eq.$\,$\ref{transRateForwardRole} does not entirely cover the contribution of the subgraph structure. Strictly, the infection rate of $\nu_S$ depends on the fraction $\eta\left(\boldsymbol{D}\right)$ of the infected neighbors which point to $\nu_S$. To elaborate this point, consider the high-order-motif shown in Fig.$\,$\ref{fig:M7M1}(a), where $\nu_S$ has five infected neighbors. This network consists of $\mathcal{M}_{7}$-triads, i.e. each infected node can potentially transmit a disease to all of it's susceptible neighbors. However, if we replace the triads with motifs of $\mathcal{M}_{1}$-type, as shown in Fig.$\,$\ref{fig:M7M1}(b), we can see that only three of $\nu_S$'s neighbors can transmit a disease to $\nu_S$. Thus, the rate at which $\nu_S$ is infected by its neighbors decreases immediately by $3/5$, compared to the network with $\mathcal{M}_{7}$-triads.
More generally, we can express this impact of the type of $\mathcal{M}_{i}$ simply by 
\begin{widetext}
\begin{equation}
\eta\left(\boldsymbol{D}\right) = \frac{\boldsymbol{\chi} \cdot \left(0,0,0,d_{ac},d_{ab},d_{bc},d_{ba},d_{cb},d_{ca},d_{ac} + d_{bc},d_{ab} + d_{cb},d_{ba} + d_{ca}\right)^{T}}{\Sigma_{i=4}^{9}\chi_{i}+ 2\Sigma_{i=10}^{12}\chi_{i}}
\end{equation} 
\end{widetext}
Hence, the forward transition rate is now given by
\begin{equation}
w^+(\boldsymbol{v}_{st}, \boldsymbol{v}_{sh}) = \beta \, p_{\nu_S}(\boldsymbol{v}_{st}, \boldsymbol{v}_{sh})\,\eta\left(\boldsymbol{D}\right)
\label{eq:transRateForwardRoleFraction}
\end{equation}
Note that the transition rate, i.e. the recovery process of $\nu_I$, is independent of the type of contact between $\nu_I$ and its neighbors. Thus, the rate $w^-(\boldsymbol{v}_{st}, \boldsymbol{v}_{sh})$ from Eq.$\,$\ref{transRateBackwardRole} does not change. Within the formalism of the NoTa-approach the fraction $\eta\left(\boldsymbol{D}\right)$ makes it strikingly simple to include the type of the TRG-building blocks. Note also, that as in Sec.$\,$\ref{sec:MeanField} we used the assumption of homogeneous rates, i.e. $\beta$ and $\gamma$ stay constant throughout the simulation.\\ 

\subsection{Numerical results}
\begin{figure}
\includegraphics[width=1.0\linewidth]{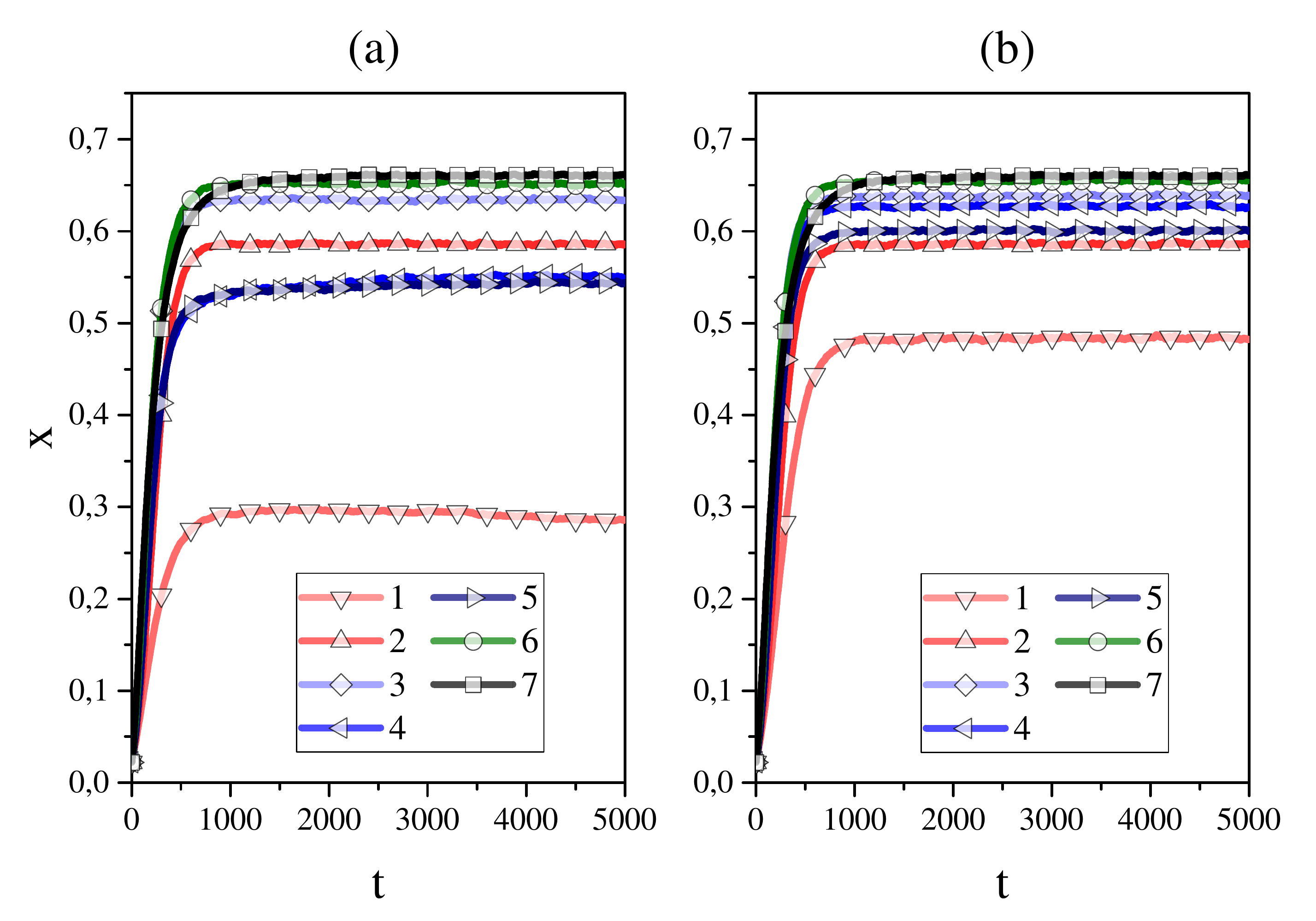}       
\caption[]{Numerical solutions to the stochastic model given by Eq.$\,$\ref{transRateBackwardRole} and \ref{eq:transRateForwardRoleFraction}. $x$ is the fraction of infected individuals, $t$ is the number of iteration steps. (a) Non-randomized $\mathcal{T}_{i}$, (b) $\mathcal{T}_{i}$ with randomized motif matrix $\boldsymbol{D}$. The results in (a) and (b) are qualitatively similar to the mean-field simulations shown in Fig.$\,$\ref{fig:nimfa60}. For $\mathcal{T}_{1}$ the number of infected individuals is clearly lower than for other TRGs. In the endemic state, the highest values of $x_{end}$ can be observed on $\mathcal{T}_{3}$, $\mathcal{T}_{6}$ and $\mathcal{T}_{7}$ and $x_{end}$ for $\mathcal{T}_{4}$ and $\mathcal{T}_{5}$ lies again between $\mathcal{T}_{2}$ and $\mathcal{T}_{1}$. Randomization leads to an increase of the values of $x_{end}$ for $\mathcal{T}_{4}$ and $\mathcal{T}_{5}$, as well as a clear increase for $\mathcal{T}_{1}$, while it doesn't change for other TRGs. Parameters: $N=49$, $T=60$, $\beta=6\cdot10^3$, $\gamma=10^3$, $x(0) = 0.02$. The maximum values of $\theta$ are: $\theta_1 = \theta_2 = 8$, $\theta_3 = \theta_4 = \theta_5 = 10$, $\theta_{6} = 12$ and $\theta_{7} = 14$ (number in the subscript corresponds to the index of $\mathcal{T}_{i}$).}   
\label{fig:stocha}    
\end{figure} 
Fig.$\,$\ref{fig:stocha} shows the evolution of the density of infected population $x$ for different $\mathcal{T}_{i}$. Note that the fraction of infected individuals is proportionate to the fraction of infected tags:
\begin{equation}
x = \frac{1}{N} \Sigma_{j}^{n_I} \left(\frac{a_{I,j} + b_{I,j} + c_{I,j}}{\kappa}\right), 
\end{equation}
where $\kappa$ is the average number of tags per node. Here, we set simply $\kappa = \frac{3T}{N}$, which eventually results in:
\begin{equation}
x = \frac{n_I}{N} = \frac{A_{I} + B_{I} + C_{I}}{3T}. 
\end{equation} 
These curves are qualitatively very similar to the curves obtained from mean-field equations shown in Fig.$\,$\ref{fig:nimfa60}.\\
Furthermore, assuming ergodicity we can simulate random network behavior by randomly sampling from different realizations of $\boldsymbol{D}_{i}$. As mentioned in Sec.$\,$\ref{sec:TRG}, the in- and out-degrees of the node do not change for the null-models. This means that $\nu$ and its neighbors can still be separated into triples. However, the tags cannot be assigned to any particular set $\left(k_{in},k_{out}\right)$, so that $\boldsymbol{D}_{i}$ can be different at every iteration step. Take the FFL from Fig.$\,$\ref{M3M6}, there are six possible configurations of assigning tags to the nodes, i.e. six realizations of $\boldsymbol{D}_{i}$, whereas for a FBL there are only 2. We simulate a randomized network by uniformly sampling from all possible configurations of $\boldsymbol{D}_{i}$. As shown in Fig.$\,$\ref{fig:stocha} (b), there are remarkable quantitative similarities to the mean-field model (Fig.$\,$\ref{fig:nimfa60} (b)). While the curves of $\mathcal{T}_{2}$, $\mathcal{T}_{6}$, $\mathcal{T}_{7}$ show no differences between the randomized and non-randomized realizations, the curves of $\mathcal{T}_{1}$ differ clearly.

\section{Conclusion}
Network analysis has become an important tool for the study of various dynamical systems, for instance financial markets, coupled oscillators or epidemics. Examining the properties of real-world networks such as community structures, hierarchical ordering or overrepresentation of specific motifs can lead to new insights on how the parameters of the system evolve in time. Supported by the discovery that certain triadic subpatterns occur in real-world data more frequently than expected from random behavior \cite{Alon}, much attention has been devoted to generating networks with an abundance of such motifs and to incorporate triangular subgraphs and triple-closure into the analysis of disease spread. However, numerous growth models assume conditionally independent dyads although this assumption might not be valid for networks with high quantities of triadic patterns \cite{Marco}.\\
In order to get a better understanding of how the latter might influence the dynamics of the Susceptible-Infected-Susceptible model, we generated ensembles of Triadic Random Graphs $\mathcal{T}_{i}$, a special network-class which uses distinct triadic motifs $\mathcal{M}_{i}$ as building blocks. We compared various global properties of TRGs which differ in the type of their motifs and found many remarkable equalities as well as differences. For instance, TRGs constructed from feed-forward-loop motifs have the same number of edges, average total degree and average clustering coefficient as TRGs constructed from feed-back-loops but they differ largely in their diameter, mean distance, assortativity and current parameter (i.e. presence of hierarchical structure).\\
Strikingly, the outcome of simulations on epidemic spread, modelled by the N-intertwined mean-field approximation (NIMFA) \cite{Mieghem2011}, shows clear difference between certain types of TRGs. By comparing the simulation results on non-randomized TRGs and the corresponding null-models, we observe that on most $\mathcal{T}_{i}$ the fraction of infected individuals evolves in time in the same way (up to small deviations). However, the abundance of feed-forward-loops decreases this fraction significantly which is mainly due to the presence of nodes with zero in- or out-degree and the occurrence of a hierarchical structure.\\
Furthermore, we developed a novel stochastic model based on assigning tags to the nodes of each $\mathcal{M}_{i}$ according to corresponding sets of $\left(k_{in},k_{out}\right)$. Instead of including whole preconstructed networks in terms of an adjacency matrix, the only topological parameters used as input are the total number of nodes $N$ and triads $T$ and the type of triadic motifs described by the motif matrix $\boldsymbol{D}$. Nevertheless, qualitatively, the simulation outcome resembles the NIMFA curves remarkably well. Hence, the motif structure has clearly an impact on the epidemic dynamics.\\
These results could improve the understanding of how certain topological patterns change the outcome of an epidemic and how dynamic parameters can be influenced by specific types of networks.

\bibliography{SIS epidemics on Triadic Random Graphs.bib}

\end{document}